\documentclass[aps,pra,twocolumn,groupedaddress,footinbib]{revtex4-2}
\usepackage{graphicx}
\usepackage{amsmath, amssymb}
\usepackage{braket}

\begin{document}

%\preprint{}

\title{Non-equilibrium mode competition in a pumped dye-filled cavity}



\author{M. Vlaho}
\email[]{vlaho@tu-berlin.de}
\affiliation{Institut f\"ur Theoretische Physik, Technische Universit\"at Berlin, Hardenbergstrasse 36, 10623 Berlin, Germany}
%\affiliation{Institut für Theoretische Physik, Technische Universität Berlin, Hardenbergstrasse 36, 10623 Berlin, Germany}

\author{A. Eckardt}
\email[]{eckardt@tu-berlin.de}
\affiliation{Institut f\"ur Theoretische Physik, Technische Universit\"at Berlin, Hardenbergstrasse 36, 10623 Berlin, Germany}
%\affiliation{Institut für Theoretische Physik, Technische Universität Berlin, Hardenbergstrasse 36, 10623 Berlin, Germany}


\date{\today}

\begin{abstract}
We consider a homogeneously pumped photon gas coupled to a dye medium and investigate how its steady state is effected when varying the pump power, the photon cavity lifetime and the cutoff frequency. We study how the interplay between pumping, loss, and dye-induced thermalization influences the selection of the cavity modes that acquire large occupation. Depending on the parameter regime, the latter can be related either to lasing of (typically multiple) modes or to equilibrium-like photon condensation in the ground mode. We calculate and explain the phase diagram of the system, with a particular emphasis on the role played by a mode repulsion that occurs in the regime of weak cavity loss.
\end{abstract}

%\keywords{}

\maketitle


\section{Introduction}
Following the realization of Bose-Einstein condensation (BEC) of exciton-polaritons, \cite{deng_condensation_2002,kasprzak_bose-einstein_2006,byrnes_excitonpolariton_2014,balili_bose-einstein_2007,plumhof_room-temperature_2014,sun_bose-einstein_2017,sun_observation_2012}, thermalization \cite{klaers_thermalization_2010} and the formation of an equilibrium-like Bose condensate of photons has been realized in various experimental setups \cite{klaers_boseeinstein_2010, marelic_experimental_2015,weill_boseeinstein_2019,rajan_photon_2016,nyman_bose-einstein_2018,walker_driven-dissipative_2018,marelic2016spatiotemporal}. The inherently grand-canonical statistics of photon BECs \cite{schmitt_observation_2014} has been studied, as well as its spatial \cite{keeling_spatial_2016} and temporal features \cite{kirton_thermalization_2015, hesten_collective_2018,hesten_non-critical_2018,schmitt_dynamics_2018}. A lot of work has also been done to clarify the delimitation of photon BECs from lasers \cite{schmitt_thermalization_2015,schmitt_bose-einstein_2016,leymann_pump-power-driven_2017,vorberg2018unified,radonjic_interplay_2018} and thermo-optic imprinting has been used to create variable potentials for coupled photon condensates.

Given the inherently driven-dissipative nature of these systems, a complex interplay between the  pump and loss processes driving the system out of equilibrium and the thermalizing influence of the environment (dye solution), emerges when tuning the various control parameters. This non-equilibrium nature of these systems makes them an excellent experimental platform for studying ordering under non-equilibrium conditions. First studies of this physics have recently been published. The case of a symmetrically pumped system and its steady-state featuring multi-mode condensates has been studied theoretically 
\cite{hesten_decondensation_2018}, predicting a complex phase diagram, where the phases are distinguished by their different combinations of macroscopically populated modes, as well as experimentally \cite{walker_driven-dissipative_2018}. Also, the case where the system is driven by an off-centered pump beam has been investigated and shown to feature a robust mechanism for controlled two-mode emission \cite{vlaho2019controlled}.

Here we consider the case of a symmetrically (homogeneously) pumped system, in which multi-mode condensates occur. We show how the limit of quasi-equilibrium photon BEC is approached via mode competition when the photon cavity lifetime is increased. Moreover, we also discuss the role played by the cutoff frequency, i.e. the ground-mode energy, in the formation of multi-mode condensates.

The remaining part of this paper is organized as follows. In section II, we introduce the model system, described in terms of rate equations for the photon mode populations $n_{i}$, and the spatially dependent fraction $f(\vec{r})$ of excited dye molecules. Then, in section III, we discuss the condition for mode selection, i.e. a mode acquiring macroscopic occupation. We also explain how this condition is connected to the locking of the chemical potential as a BEC condition in the equilibrium limit (Sec. III A). We give a general condition for the threshold pump rate fo a mode selection, which reduces to very simple expression in the case of first selection (Sec. III B). Before turning to discuss our results, we list all the parameter values corresponding to the relevant experiments and used in the numerical simulations (Sec. III C). In the following section (IV), we investigate the dependence of mode selection on the so-called thermalization parameter, which is effectively a dimensionless photon cavity lifetime. Here we observe and discuss the effects of mode repulsion and deselection (loss of macroscopic occupation) \cite{hesten_decondensation_2018}, before presenting a phase diagram in a parameter plane spanned by the pump rate and the thermalization parameter (Sec. V). We explain the various phase boundaries and how the approach towards a quasi-equilibrium ground-mode condensate manifests in the phase diagram. Moreover, we point out discrepancies with respect to the previously computed phase diagram of Ref.~\cite{hesten_decondensation_2018}. In the final section (VI) we study how the mode selection is effected when tuning the cutoff-frequency.

\section{System and model}
\begin{figure}[!htbp]
	\includegraphics{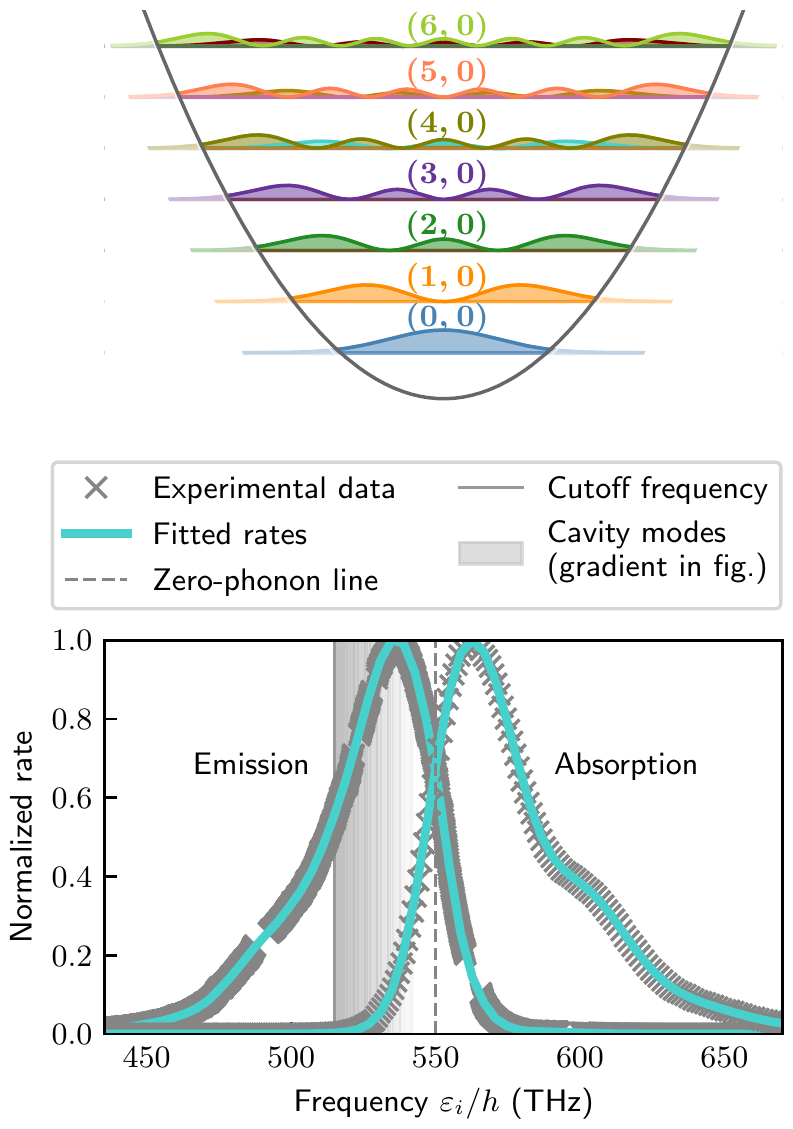}
	\caption{\label{fig:modes} Upper panel: Photon mode densities $|\psi_{n_x, n_y}(x)|^{2}$ projected onto the $x$ axis. Modes $(n_x,0)$, with nodes only along $x$ direction, are labeled. Lower panel: Fitted absorption and emission rates (solid lines) vs. frequency $ \omega / 2 \pi $. The rates are fitted to experimental data (crosses) \cite{keeling_spatial_2016, nyman_absorption_2017}. The frequency range of the relevant cavity modes is indicated by the shaded gray area with a sharp cutoff at $ \omega_c = \varepsilon_0 / \hbar $.}
\end{figure}

We describe the system in terms of semiclassical equations of motion \cite{kirton_nonequilibrium_2013, keeling_spatial_2016, hesten_decondensation_2018, vlaho2019controlled} for the photon mode populations $n_{i}$, and the fraction $f(\vec{r})$ of excited dye molecules at position $ \vec{r} $ in the two-dimensional space given by the directions vertical to the optical axis of the cavity,

\begin{eqnarray}
\label{dyn1} \dot{n}_{i} &=&	-\kappa n_{i} + (n_{i}+1) R_{\downarrow}^{i} \rho \, G_i 
-n_{i}R_{\uparrow}^{i} \rho \, (1-G_i), \\
\label{dyn2} \dot{f}(\vec{r})&=& [1-f(\vec{r})]P+ \sum_{i}R_{\uparrow}^{i}|\psi_{i}(\vec{r})|^{2}n_{i}) 
\nonumber\\
&& -\,f(\vec{r})[\Gamma + \sum_{i}R_{\downarrow}^{i}|\psi_{i}(\vec{r})|^{2}(n_{i}+1)].
\end{eqnarray}
Here $\rho$ is the density of the dye molecules, $\Gamma$ is the rate of spontaneous losses into non-cavity modes and $\kappa$ the photon loss rate. The transverse photonic modes $\psi_{i}(\vec{r})$ (upper panel of Fig.~\ref{fig:modes}) are the eigenfunctions of the two-dimensional harmonic trap imposed by the spherically curved mirrors. They are characterized by a pair of harmonic oscillator quantum numbers in $ x $ and $ y $ direction, $ i = (\nu_x,\nu_y) $, and have energies $E_i=\hbar [\Omega_x(\nu_x+1/2)+ \Omega_y(\nu_y+1/2)]$  with oscillator frequencies $ \Omega_x $ and $ \Omega_y $. In the following, we assume  $ \Omega_x $ and $ \Omega_y $ to be almost identical, $ \Omega_x \equiv \Omega $ and $ \Omega_y=0.99 \,\Omega $. This slight anisotropy of the trap is a realistic assumption and is required for eliminating the coherent mixing of otherwise degenerate modes $i$. The harmonic oscillator length $d$ associated with $ \Omega $ is used as a natural unit of length. The total energy of the cavity mode $i$ is then given by $ \varepsilon_i = E_i + \hbar \omega_L$, where $ \omega_L $ is the frequency of the relevant longitudinal mode.
The gain of mode $i$ is quantified by its overlap with the fraction $ f(\vec{r}) $ of excited dye molecules,
\begin{equation}\label{gain_def}
G_i[f(\vec{r})] = \int |\psi_{i}(\vec{r})|^{2} f(\vec{r})\, d\vec{r}.
\end{equation}
The dye solution is characterized by the absorption and emission rates, $R_{\uparrow}^{i}$ and $R_{\downarrow}^{i}$, which satisfy the Kennard-Stepanov law \cite{mccumber1964einstein,klaers_thermalization_2010,moroshkin2014kennard}
\begin{equation}\label{KS_law}
R_{\downarrow}^{i}/R_{\uparrow}^{i}= C e^{-\beta (\varepsilon_i-\hbar \omega_{z})},
\end{equation}
that can enable the photon gas to thermalize. Here, $ \omega_{z} $ denotes the zero-phonon frequency of the dye and $ C $ is a frequency-independent proportionality constant. The pump rate $P$ is considered to be spatially constant \footnote{The pump can also be modeled as a very wide Gaussian beam, to more accurately correspond to the experiment. This introduces only minor quantitative changes into the work presented here, without effecting any of the qualitative results.}.

\section{Mode selection}
We can write the steady state of Eq.~\eqref{dyn1} in the following form
\begin{equation}\label{steady}
n_i =\left(\frac{R_{\uparrow}^{i}}{R_{\downarrow}^{i}}\frac{(1-G_i)}{G_i}-1+ \dfrac{\kappa}{R_{\downarrow}^{i} \rho G_i}\right)^{-1}.
\end{equation}

When a mode $ i $ becomes macroscopically occupied (``Bose selected'' \cite{vorberg_generalized_2013, leymann_pump-power-driven_2017, vorberg2015nonequilibrium, vorberg2018unified}), the contribution of spontaneous emission to this macroscopic population $ n_i $ becomes negligible ($ n_i+1 \approx n_i $). This allows us to find a sharply defined threshold value $ G_i^{th} $ of the gain at which the ``selection'' happens. It is obtained by setting the term in the brackets to zero (corresponding to a divergent occupation). We get
\begin{equation}\label{thres}
G_i^{th} = \frac{R_{\uparrow}^{i}+\kappa / \rho}{R_{\uparrow}^{i}+ R_{\downarrow}^{i}} = \frac{1+ R_{\uparrow}^0 /(R_{\uparrow}^i \xi)}{1+ C e^{-\beta (\varepsilon_i-\hbar \omega_{z})}},
\end{equation}
where we have isolated the thermalization parameter \cite{keeling_spatial_2016, hesten_decondensation_2018}
\begin{equation}\label{xi_def}
\xi=R_{\uparrow}^{0}\rho/\kappa
\end{equation}
as a dimensionless measure of the coupling between the photons and the dye relative to the cavity loss. Once a mode is selected, the gain $G_i$ is clamped \cite{siegman_lasers_1986} close to the threshold value $G_i^{th}$.

\subsection{Equilibrium limit}
In the case of equilibrium Bose-Einstein condensation, this ``divergent'' (macroscopic) occupation in the ground mode happens when the chemical potential approaches the value of ground mode energy. This locking of the chemical potential can be shown to be equivalent to the above defined clamping of the gain $ G_i $ in the limit $ \xi \rightarrow \infty$ for the case of homogeneous excitation field $f(\vec{r}) = const. \equiv f$. Equation \eqref{steady} then reduces to the Bose-Einstein distribution
\begin{equation}\label{BE}
n_i =\left( e^{\beta (\varepsilon_i - \mu)}-1\right)^{-1},
\end{equation}
where we have used the Kennard-Stepanov law [Eq. \eqref{KS_law}] and introduced the chemical potential $ \mu $ of the photon gas given by \cite{klaers2012statistical}
\begin{equation}\label{chem}
e^{\beta \mu} = C e^{\beta \hbar \omega_{z}} f/(1-f),
\end{equation}
where $ f/(1-f) $ is now a spatially homogeneous ratio of the number of excited and ground-state dye molecules. When $ \mu \rightarrow \varepsilon_0 $ (onset of BEC), Eq. \eqref{chem} becomes $C e^{-\beta (\varepsilon_0-\hbar \omega_{z})}=(1-f)/f$. It follows that $f = G_0^{th}=1/(1+ C e^{-\beta (\varepsilon_0-\hbar \omega_{z})})$, which is exactly the selection threshold condition for the ground mode given by Eq.~\eqref{thres} when the photon cavity lifetime $ 1/\kappa \propto \xi \rightarrow \infty $. Therefore, in the equilibrium limit, which does not require pumping in order to stabilize the average photon number, the locking of the chemical potential is equivalent to the clamping of the gain.

\subsection{Threshold pump rate}
The general condition for the selection threshold pump rate $ P^{th}_{i} $ of mode $i$ can be obtained by inserting Eq.~\eqref{dyn2} into the definition of the gain \eqref{gain_def} and setting it equal to $ G_i^{th} $. We get \cite{hesten_decondensation_2018}
\begin{equation}\label{P-th-cond-general}
G_{i}^{th} = \int d\vec{r} \,|\psi_{i}(\vec{r})|^{2}\,\frac{P+\sum_{j \in S} R_{\uparrow}^{j}\,|\psi_{j}(\vec{r})|^{2}\, n_{j}}{\Gamma+P+\sum_{j \in S} (R_{\uparrow}^{j}+R_{\downarrow}^{j})\,|\psi_{j}(\vec{r})|^{2}\, n_{j}},
\end{equation}
where the sums in the integrand are over all of the modes which have already been selected. Again, the sharpness of transitions allows us to omit the contributions of spontaneous emission to the mean occupations, which are negligible once a mode is selected.
\begin{figure}[!htbp]
	\vspace{-0.7cm}
	\includegraphics[scale=1]{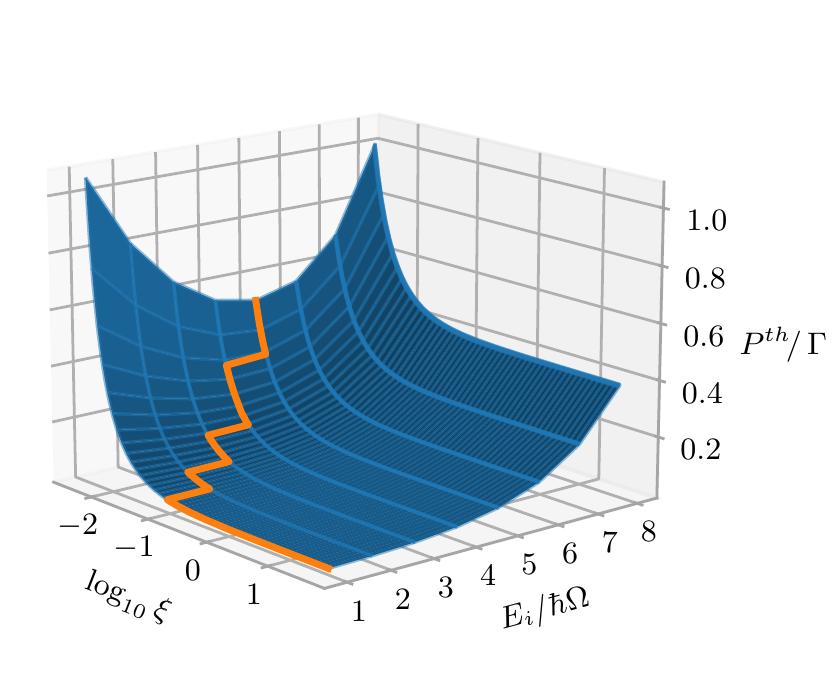}
	\vspace{-0.3cm}
	\caption{Threshold pump rate $P_{th}$ of the first selection as a function of the thermalization parameter $\xi$ and the energy $E_i$ of the modes.}
	\label{fig:surface}
\end{figure}

This equation would allow us, in principle, to iteratively determine each selected mode and the corresponding threshold pump rate, if the populations of all the already selected modes are known as a function of the pump rate $ P $. Namely, at each value of $\xi$, the selected mode is the one for which the Eq.~\eqref{P-th-cond-general} holds for the lowest value of $ P $.
However, an analytic expression can only be obtained for the first selection. Here, the approximate value of the threshold $ P^{th}_{i} $ can be obtained by setting all $ n_j $ to zero (i.e. neglecting the coupling to the still weakly occupied photonic modes). Equation \eqref{P-th-cond-general} then reduces to
\begin{equation}\label{}
G_{i}^{th} = \frac{P}{\Gamma+P}.
\end{equation}
Solving this equation for $P$ and inserting the condition \eqref{thres}, we get the first-selection threshold pump rate $P^{th}_{i}$ as a function of the thermalization parameter $\xi$
\begin{equation}\label{P_th}
P^{th}_{i} = \frac{G_{i}^{th}}{1-G_{i}^{th}}\Gamma=\frac{R_{\uparrow}^{i}+R_{\uparrow}^{0}/\xi}{R_{\downarrow}^{i}-R_{\uparrow}^{0}/\xi} \ \Gamma.
\end{equation}
It follows that the first-selected mode is the one with the lowest threshold gain $G^{th}_{i}(\xi)$. In Fig.~\ref{fig:surface} the threshold pump rate $P^{th}_i$ is shown as a function of $\xi$ and the mode energy $ E_i $ (blue surface). The orange line on this surface follows the minimal $P^{th}_{i}$ at each value of $ \xi $. We see that only for very small values of the thermalization parameter, $ \xi \lesssim 0.1 $, this is not the ground mode. This follows from Eq. \eqref{thres} when taking into account the shapes of the absorption and emission spectra (see the lower panel of Fig.~\ref{fig:modes}). In particular, by looking at the second equation we see that the $\xi$-independent denominator always favors the ground mode which has the largest Boltzmann factor, whereas the nominator contains the relative absorption rate of mode $i$, $ R_{\uparrow}^{i}/R_{\uparrow}^{0} $, modulated by the thermalization parameter. When the latter is sufficiently large, $ \xi \gtrsim 0.1 $, this term becomes negligible compared to 1, and the selected mode is determined solely by the ground-mode favoring Boltzmann factor. On the other hand, in the high loss regime, $ \xi \lesssim 0.1 $, excited modes with higher relative absorption rate can ``win out", i.e. have the smallest threshold gain.
\begin{figure*}[!htbp]
	\includegraphics[scale=1]{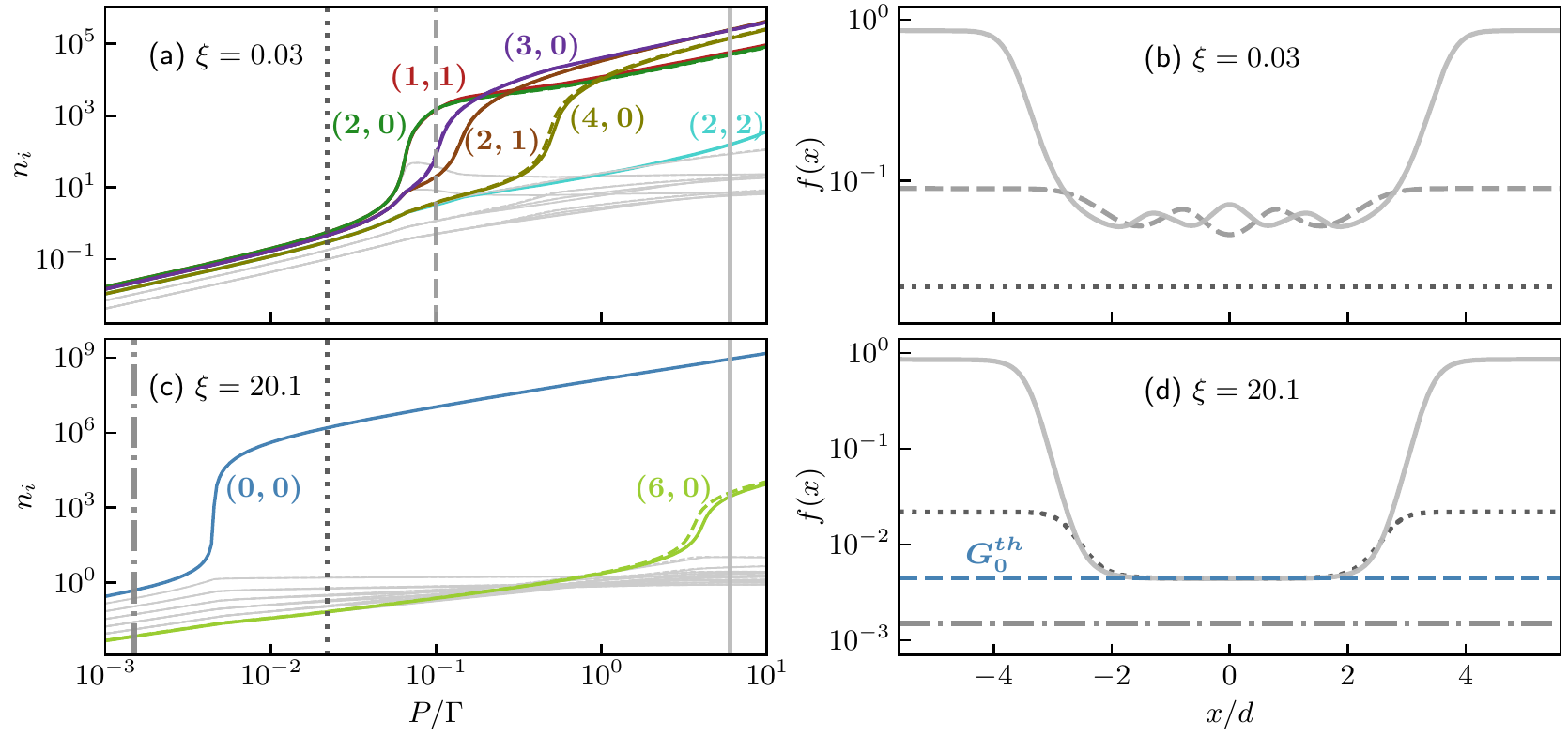}
	\caption{The left panel shows the mode populations $ n_i $ as functions of pump rate $ P $ for two values of the thermalization parameter, $ \xi = 0.03 $ (a) and $ \xi = 20.1 $ (c). Only one of the modes in an almost identically behaving symmetric pair is labeled. The right panel, (b) and (c), shows the corresponding spatial distributions of excited dye molecules $ f(x) $ for the chosen values of $P$ (vertical lines in the left panel).}
	\label{fig:homog1}
\end{figure*}

\subsection{System parameters}
For our numerical calculation we use the parameter values which correspond to the experiments of Refs.~\cite{schmitt_thermalization_2015, dung_variable_2017}. Namely, we choose a slightly anisotropic harmonic trap, as defined above, with $\Omega/2\pi=4~\mathrm{THz}$. The ground-mode (cutoff) frequency is set to $\omega_c = \varepsilon_0/\hbar = 2 \pi \cdot 515~\mathrm{THz}$. From the measured absorption and fluorescence spectra of the Rhodamine 6G dye \cite{keeling_spatial_2016}, we obtain the corresponding rates $R_{\uparrow, \downarrow}^{i}$ as fitted functions of the frequency $E_i/\hbar$ \cite{keeling_spatial_2016, nyman_absorption_2017}, as shown in the lower panel of Fig.~\ref{fig:modes}. The values of the absorption and emission rates across the whole frequency range are then determined by setting the absorption rate of the ground mode to $R_{\uparrow}^{0}/ d^2=1~\mathrm{kHz}$. The density of dye molecules is set to $ \rho = 10^8 / d^2 $ and the thermalization parameter $\xi$ lies between $0.01$ and $100$, where $\xi = 25$ corresponds to the mean experimental value of the photon loss rate $ \kappa \approx 4 $~GHz \cite{schmitt_thermalization_2015}. The rate of spontaneous losses into non-cavity modes is set to $\Gamma = 0.2~\mathrm{GHz}$.

\section{Tuning the photon cavity lifetime and mode repulsion}
Let us now discuss how the selection of modes is influenced by the thermalisation parameter $ \xi $, or equivalently the photon cavity lifetime $ 1/\kappa $. Numerically obtained photon mode populations as functions of the pump rate are shown in Fig.~\ref{fig:homog1} for two values of the thermalization parameter $\xi$. The colors correspond to the modes as shown in Fig.~\ref{fig:modes} and only modes with varying shapes are shown with different colors, e.g. symmetric pairs like (2,1) and (1,2) are depicted with the same color (brown), but a different linestyle (solid vs. dashed). These same colors are used consistently in all the figures of this paper.
As expected from Fig.~\ref{fig:surface}, for the low value $ \xi=0.03 $ (Fig.~\ref{fig:homog1}a), multiple quasi-degenerate excited modes (2,0), (0,2) and (1,1), colored green and red, respectively, are selected at practically identical threshold pump rate. This is followed by further selection of modes with higher energy, while the ground mode (0,0) remains unselected. 

When $ \xi $ is sufficiently large, as shown in Fig.~\ref{fig:homog1}c, the ground mode is the first one to get selected at a much lower value of $P$. However, as $ P $ is increased further, eventually also modes (6,0) and (0,6) get selected, as opposed to the energetically favorable selection of the first excited modes (1,0) and (0,1), which according to Eq.~\eqref{thres} possess a lower threshold gain $ G_i^{th} $.

In order to explain this, we plot the fraction of excited dye molecules $ f(x)\equiv f(x,0) $ in Fig.~\ref{fig:homog1}d for three chosen values of P (indicated by vertical lines of the same style in the left panel). We see that once $ P $ is increased above the first selection threshold, the shape of $ f(x) $ reflects the clamping of the gain in the central region, which overlaps with the selected ground-state mode (Fig.~\ref{fig:homog1}d). This clamping in the center of the trap then suppresses the selection of further low-energy modes, whose wave functions have a large weight in the trap center. This mechanism of mode repulsion explains why after the ground mode, the next modes to be selected posses six excitation quanta. Likewise, Fig.~\ref{fig:homog1}b shows that after the first selection $ f(x) $ reflects the shape of the selected excited modes for the scenario of Fig.~\ref{fig:homog1}a.
\begin{figure}[!htbp]
	\includegraphics[scale=1]{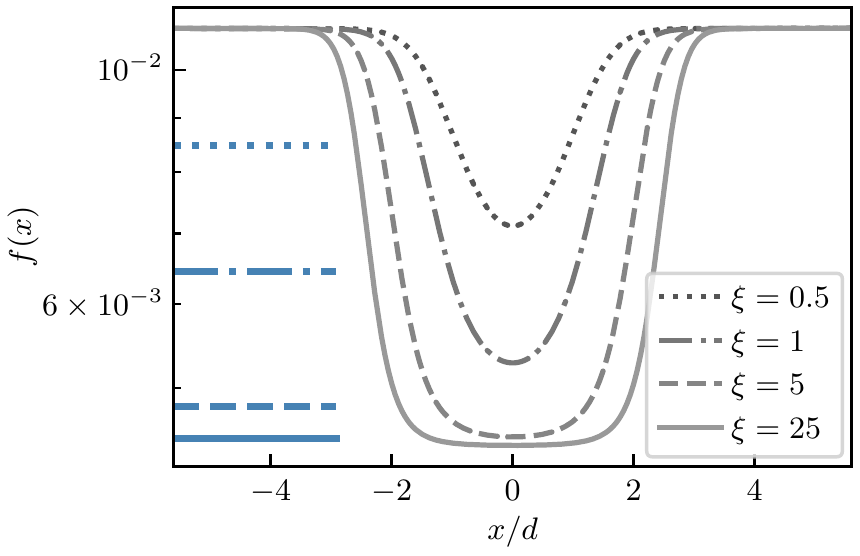}
	\caption{Spatial distributions of excited dye molecules $ f(x) $ close above the first selection threshold $ P^{th} $ for 4 values of the thermalization parameter $\xi$.}
	\label{fig:f_comp}
\end{figure}

To see more clearly how the shape of $ f(x) $ and with that the mode repulsion changes with $ \xi $, we plot its value at $ P = 0.02 $ for 4 values of $ \xi $ (Fig.~\ref{fig:f_comp}). This pump rate is slightly above the first selection threshold $ P^{th} $ and only the ground mode is selected in each case. It follows from Eqs.~\eqref{thres} and \eqref{P_th} that a higher value of $ \xi $ lowers the first selection threshold and the corresponding gain gets clamped at a lower value $ G^{th} $, indicated by horizontal blue lines for each $ \xi $ in Fig.~\ref{fig:f_comp}. We see that with increase of $ \xi $ the excited dye molecules are clamped both at a lower value and in a wider spatial region, therefore becoming progressively more inaccessible to modes close in energy to the ground mode. In other words, when the gain is clamped at lower values, not enough molecules can be excited in the region of overlap between the ground mode and following excited modes. We can say that successive modes (those with largest overlap) “compete" for gain in the same spatial domain and block each other from being selected together in the regime of higher P. When the thermalization parameter is increased even further to $ \xi = 20.1 $ (Fig.~\ref{fig:homog1}d) $ f(x) $ is locked to an even lower threshold value of the gain $ G_{0}^{th} $ (dashed blue line) in an even wider middle region. In this way, when increasing $ \xi $, a quasi-equilibrium steady state is approached, where the gain clamping is equivalent to locking of the chemical potential. The fact that $ f(x) $ is free to increase with the pump power in the outer spatial region until gain saturation is reached (no more dye molecules available to excite) reflects the non-equilibrium nature of the system.
\begin{figure}[!htbp]
	\includegraphics[scale=1]{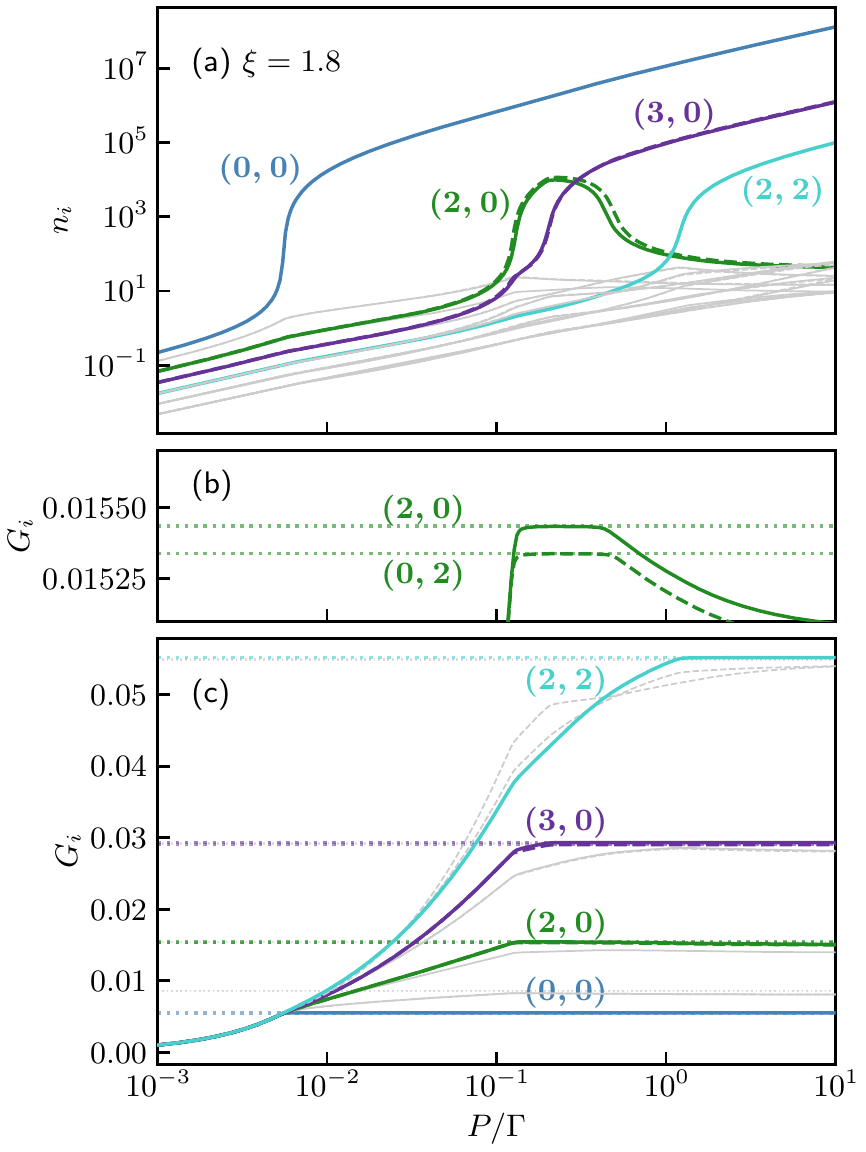}
	\caption{Population $ n_i $ (a) and gain $ G_i $ (c) of modes $i$ vs. pump rate $ P $ for $\xi = 1.8 $. Only one of the modes in an almost identically behaving symmetric pair is labeled. Dotted horizontal lines indicate threshold values $ G_i^{th} $ to which the gain is clamped at selection. The middle panel (b) shows the zoomed-in gain of modes (2,0) and (0,2), which are deselected when the clamped gain starts to drop below $ G_{(2,0)}^{th} $ and $ G_{(0,2)}^{th} $, respectively.}
	\label{fig:homog2}
\end{figure}

Another observation that we can make from Fig.~\ref{fig:homog1} is that, while the modes (6,0) and (0,6) are selected in high $ P $ regime, all the other modes with the \emph{same} energy (like (3,3), (4,2), (5,1) etc.) remain unselected due to the key influence of the dye excitation profile on the behavior of modes. This is another indicator of the non-equilibrium nature of this state (even though at lower $ P $ where only (0,0) is selected, this state can hardly be distinguished from the equilibrium photon BEC (of a finite system), as will be discussed below in more detail).
\begin{figure*}[!htbp]
	\includegraphics[scale=1]{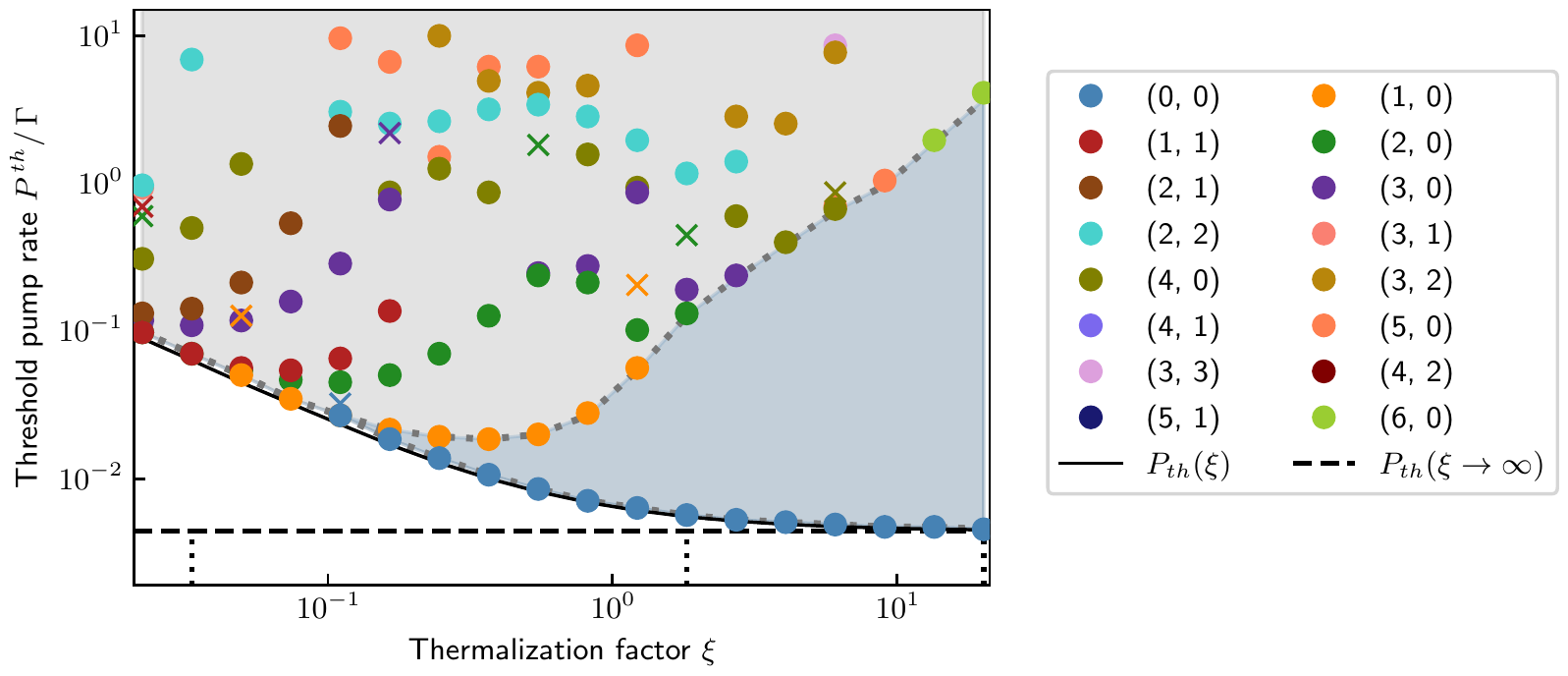}
	\caption{\label{fig:pd} Phase diagram showing three main regions. The white region has no selected modes, only the ground mode is selected in the blue region, and there are multiple selections in the gray one. Colored dots (crosses) are numerical points indicating selections (deselections) of the corresponding modes. The dotted line is the lower (upper) phase boundary, interpolating between the points of first (second) selection. The pump rate at which the two phase boundaries meet is the minimal $ P $ for which the ground mode is selected. Below this pump rate (``lasing phase"), the phase boundaries are indistinguishable, because multiple quasi-degenerate higher energy modes are selected at almost exactly the same $ P $. The analytical result for the first selection threshold (Eq.~\eqref{P_th}), is shown with a solid black line. In the high $ \xi $ regime, this phase boundary approaches the dashed horizontal line, showing the high thermalization limit of Eq.~\eqref{P_th} $ P^{th} (\xi \rightarrow \infty) $. The three dotted vertical lines mark the cuts through the phase diagram shown in Fig~\ref{fig:homog1} and Fig~\ref{fig:homog2}.}
\end{figure*}

In Fig.~\ref{fig:homog2}, at an intermediate value of the thermalization parameter $\xi=1.8$, we can also observe the phenomenon of ``deselection". Namely, we can see that the green-colored mode pair (2,0) and (0,2) gets deselected, as the purple-colored one, (3,0) and (0,3) is selected (Fig.~\ref{fig:homog2}a) [henceforth for brevity, mode pairs $ \{(i,j), \,(j,i), \, i \neq  j\} $ are denoted simply as ``mode pair $ (i,j) $’’]. This decondensation of photonic modes was already discussed in Ref.~\cite{hesten_decondensation_2018}. This is another manifestation of the above-mentioned competition between successive modes which have a large spatial overlap. It happens when not enough dye molecules are pumped to their excited states in the combined region of mode density for both mode pairs to stay selected. Once, after the ground mode, the mode pair (2,0) is selected, $ f(\vec{r}) $ can only increase in a very restricted way, such that it stays clamped close to both $ G_{(0,0)} $ and $ G_{(2,0)} $ (dotted lines in Fig.~\ref{fig:homog2}c). This restriction still allows for the selection of the mode pair (3,0). However, the more particles this third selected pair acquires, the more attractive it becomes for further photons due to bosonic enhancement (i.e. stimulated emission). This non-linear effect leads to a competition with the energetically slightly favored (2,0) modes, which eventually causes the decondensation of the latter. This is accompanied by the ``declamping" of its gain, which is better visible in Fig.~\ref{fig:homog2}b showing the zoom-in around $ G_{(2,0)}^{th} $.

\section{Phase diagram}
After having discussed the role of mode repulsion for the selection of excited cavity modes, let us now compute the phase diagram of the system in the parameter plane spanned by the pump rate and the thermalization parameter. Figure \ref{fig:pd} shows a phase diagram where the various phases are characterized by which modes are selected. There are no macroscopically occupied modes in the white region. Only the ground mode is selected in the blue region, whereas in the gray one, there are multiple selected modes. Colored dots (crosses) are numerical points indicating selections (deselections) of the corresponding modes. The lower and upper phase boundaries (dotted lines) are obtained by interpolating between the points of first (second) selection. Below the value of $ \xi $ at which both boundaries meet, the mode which become selected first is not the ground state anymore. Below this point, the two phase boundaries are indistinguishable, because multiple quasi-degenerate higher energy modes are selected at almost exactly the same $ P $. The solid black line indicates the analytical result for the first selection threshold, given by Eq.~\eqref{P_th} and shown with the orange line in Fig.~\ref{fig:surface}. It closely matches the numerical result, especially in the high $ \xi $ regime, where it approaches the high thermalization limit of Eq.~\eqref{P_th} $ P^{th} (\xi \rightarrow \infty)/\Gamma = R_{\uparrow}^{0}/R_{\downarrow}^{0}$ (dashed horizontal line).

We note that our phase diagram differs from the one obtained in Ref.~\cite{hesten_decondensation_2018}. Namely, in agreement with the analytical prediction \eqref{P_th}, we find that the threshold pump rate for the first selection process decreases as a function of the thermalization parameter, while it increases in Ref~\cite{hesten_decondensation_2018}.

In the regime of low $ \xi $, where the photon cavity lifetime is too short for photons to effectively thermalize, there are multiple high energy modes selected closely together instead of the ground mode, and these transitions to macroscopic occupation represent the limit where the operation of the system would typically be considered as that of a laser. Given that drive and thermalization are both present in this system, Bose condensation cannot be sharply distinguished from lasing. Nevertheless, this phase diagram still clearly shows the trend of going from the lasing limit towards the BEC limit as the thermalization parameter is ramped up. 

The second phase boundary separates the blue region with only the ground state selected from the gray one, where also excited modes have acquired a large occupation. With increasing $\xi$, the separation between the $ P^{th}_0$ of the ground mode selection and the $ P^{th}_j$ of the next selected mode $ j $ increases together with its energy $ E_j $, due to mode competition explained above. In this way, a limit of quasi-equilibrium photon BEC is approached for large $\xi$ and pump powers well below the second selection threshold.

\begin{figure}[!htbp]
	\includegraphics[scale=1]{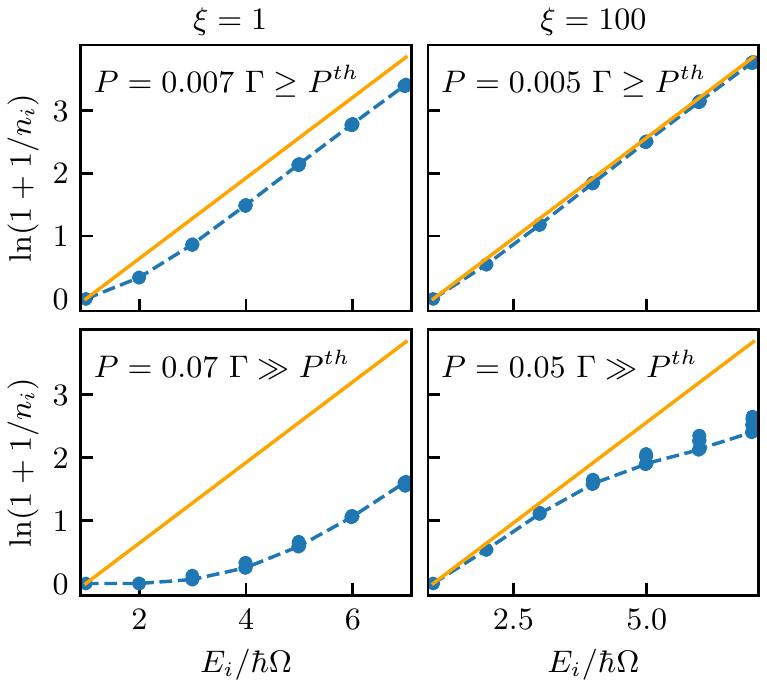}
	\caption{Numerical mode population $ n_i $ (blue dots) vs mode energy $ E_i $ compared to the thermal distribution (orange straight line) below, close above and far above the first selection threshold $ P^{th} $ when $\xi = 1 \Rightarrow P^{th} = 0.0065$ (left panels) and $ \xi = 100 \Rightarrow P^{th} = 0.0044 $ (right panels).}
	\label{fig:therm}
\end{figure}

To support this claim, in Fig.~\ref{fig:therm} we compare the Bose-Einstein (BE) distribution $ \ln(1+1/n_i) = \beta \varepsilon_i - \mu$ (straight orange line) with the distribution of numerically obtained mode populations $ n_i $ (blue dots), for $ P $ close  and far above the first selection threshold $ P^{th}$. Left panel corresponds to $ \xi=1 $, and the right one to $ \xi=100 $, for which only the ground mode is selected. As expected, for higher $ \xi $, the match between the numerical points and the thermal distribution is better, particularly when $ P $ is only slightly above the threshold, $ P \gtrsim P^{th} $. Here the small deviations are only due the fact that the absorption and emission rates fitted to measured data (Fig.~\ref{fig:modes}) do not satisfy the Kennard-Stepanov relation (Eq.~\eqref{KS_law}) exactly, but only to a very good approximation. As $ P $ is raised significantly above $ P^{th} $, the numerically obtained populations start to deviate from the BE distribution, particularly those of higher modes. This is expected, even though at the very high value of $ \xi $, no other mode is selected at $ P \gg P^{th} $ except the ground mode. The reason for this is that outside of the increasingly wide central region of the trap where the gain is clamped, $f(x)$ can still increase with $ P $. Therefore, the occupation of modes with a high density there (higher energy modes) can increase as well, moving away from the thermal distribution.

\section{Tuning the cutoff frequency}
Let us finally discuss how the physics of the mode selection changes, when considering a variation of the cutoff frequency $ \omega_c $ (or equivalently, the detuning from the zero-phonon line). The cutoff frequency corresponds to the ground-mode energy and can be tuned experimentally by varying the longitudinal frequency $\omega_L$ via the cavity length. It determines the absorption and emission rates, thus effecting also the degree of thermalization.  This has been studied experimentally for the case of a continuous wave (CW) \cite{klaers_thermalization_2010} and pulsed laser pump \cite{schmitt_thermalization_2015}, as well as in theoretical work \cite{kirton_thermalization_2015}.
\begin{figure}[!htbp]
	%\vspace{0.5cm}
	\includegraphics[scale=1]{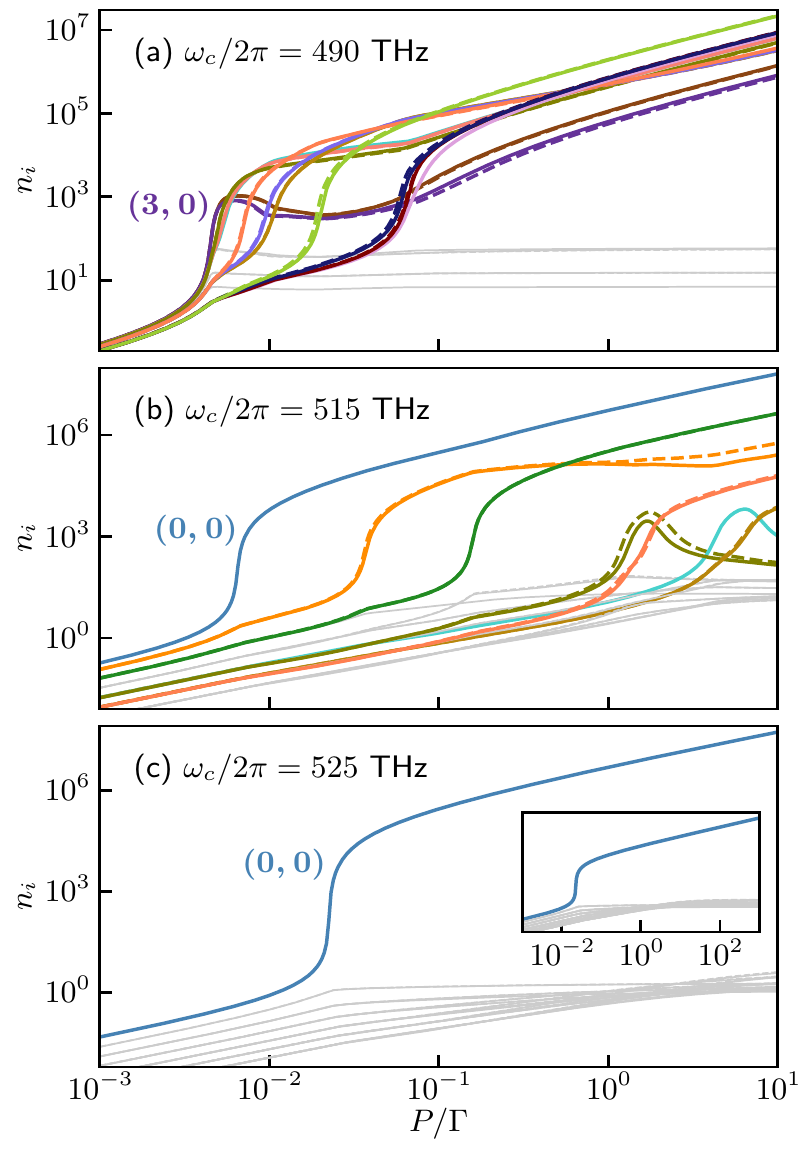}
	\caption{\label{fig:det} Population $ n_i $ of modes $i$ vs. pump rate $ P $ for $ \xi = 1 $. The cutoff frequency is $ \omega_c /2\pi= 490$ THz (a), $ \omega_c /2\pi= 515$ THz (b) and $ \omega_c /2\pi= 525$ THz (c). The inset of panel (c) contains the same result extended to high $ P $ regime, showing that only the ground mode is selected before the gain is saturated.}
\end{figure}

In Fig.~\ref{fig:det} we show the numerical results for the photon populations $ n_i $ for three different values of the cutoff frequency, while keeping the thermalization parameter fixed at $ \xi = 1 $. Changing the cutoff frequency, i.e. the ground-mode energy, corresponds to shifting the frequency range of cavity modes (sketched by the gray shaded area in the lower panel of Fig.~\ref{fig:modes}) to the left or right. This effects the threshold gain $ G_{i}^{th} $ of each mode (Eq.~\eqref{thres}).
For sufficiently low cutoff frequency $ \omega_c / 2\pi= 490$ (Fig.~\ref{fig:det}a), the absorption and emission rates no longer satisfy the Kennard-Stepanov law. Instead of the ground mode, now an excited mode pair (3,0) has the lowest value of $ G_{i}^{th} $ and it is selected first, closely followed by additional excited modes, while the ground mode remains unselected. In the case of higher $ \omega_c / 2\pi= 515$ THz (same value used for all the rest of our results), the ground mode is selected first, followed by the selection of several excited modes, as seen in Fig.~\ref{fig:det}b. Compared to Fig.~\ref{fig:det}a, the threshold pump rate $ P^{th} $ of the first selection has increased. When the cutoff frequency is shifted even further to $ \omega_c / 2\pi= 525$ THz, only the ground mode is selected before the gain is saturated (Fig.~\ref{fig:det}c) and no further selections can occur. This is shown in the inset, where the $ P $ axis is extended by 2 orders of magnitude. In this case the Kennard-Stepanov law still holds, but the corresponding $P^{th}$ and $G^{th}$ are now the highest. It should be pointed out that in the actual experiment, as the cutoff frequency is varied, the photon cavity lifetime also varies significantly \cite{schmitt_thermalization_2015}.  In the case of here chosen $\omega_c$ values, this means that $ \xi $ should increase with $\omega_c$ \footnote{Alternatively, we could consider (a), (b) and (c) to correspond to different cavities, which have the same photon loss rate $ \kappa $ at the given cutoff frequencies.}. However, this would only enhance the effect of increase in photon thermalization, observed from (a) to (c) panel in Fig.~\ref{fig:det}.

\section{Conclusions}
We have studied how the variation of the photon cavity lifetime $ 1/\kappa $ and the cutoff frequency $\omega_c$ effects the steady state of a homogeneously pumped photon gas coupled to a dye medium.
We have shown how, through the effect of mode competition (governed by the dye excitation profiles), the equilibrium-like ground-mode condensation emerges from the steady state of the system. Namely, we found that increasing the thermalization parameter $ \xi \propto 1/\kappa $ produces a form of mode repulsion, in the sense that the ground-mode selection is followed by a selection of modes with increasing number of excitation quanta. This is explained as a consequence of how the dye excitation profile $ f(\vec{r}) $ at pump powers above which the ground-mode is selected, changes with $ \xi $. We produced a phase diagram of the system in the space of 2 parameters, the pump power and the thermalization parameter and noted how it differs from the one in Ref.~\cite{hesten_decondensation_2018}.

We also looked at the effect of varying cutoff frequency $ \omega_c $ on the selection of modes, and found that, in agreement with previous work \cite{klaers_thermalization_2010, schmitt_thermalization_2015, kirton_thermalization_2015}, below a certain value of $ \omega_c $, the photons are unable to effectively thermalize, resulting in the closely spaced selections of many excited modes, as opposed to the ground mode. On the other hand, the cutoff frequency can be increased above the value used in the rest of this work, while keeping the ratio of emission and absorption rates still to a good approximation proportional to the Boltzmann factor (i.e. the Kennard-Stepanov law still holds). We show that in this case only the ground mode is selected before the gain saturates.
\\

\begin{acknowledgments}
We acknowledge the support from the Deutsche Forschungsgemeinschaft (DFG) via the Research Unit FOR 2414 (under Project No. 277974659).
\end{acknowledgments}

\bibliography{ref}

\begin{thebibliography}{39}%
\makeatletter
\providecommand \@ifxundefined [1]{%
 \@ifx{#1\undefined}
}%
\providecommand \@ifnum [1]{%
 \ifnum #1\expandafter \@firstoftwo
 \else \expandafter \@secondoftwo
 \fi
}%
\providecommand \@ifx [1]{%
 \ifx #1\expandafter \@firstoftwo
 \else \expandafter \@secondoftwo
 \fi
}%
\providecommand \natexlab [1]{#1}%
\providecommand \enquote  [1]{``#1''}%
\providecommand \bibnamefont  [1]{#1}%
\providecommand \bibfnamefont [1]{#1}%
\providecommand \citenamefont [1]{#1}%
\providecommand \href@noop [0]{\@secondoftwo}%
\providecommand \href [0]{\begingroup \@sanitize@url \@href}%
\providecommand \@href[1]{\@@startlink{#1}\@@href}%
\providecommand \@@href[1]{\endgroup#1\@@endlink}%
\providecommand \@sanitize@url [0]{\catcode `\\12\catcode `\$12\catcode
  `\&12\catcode `\#12\catcode `\^12\catcode `\_12\catcode `\%12\relax}%
\providecommand \@@startlink[1]{}%
\providecommand \@@endlink[0]{}%
\providecommand \url  [0]{\begingroup\@sanitize@url \@url }%
\providecommand \@url [1]{\endgroup\@href {#1}{\urlprefix }}%
\providecommand \urlprefix  [0]{URL }%
\providecommand \Eprint [0]{\href }%
\providecommand \doibase [0]{https://doi.org/}%
\providecommand \selectlanguage [0]{\@gobble}%
\providecommand \bibinfo  [0]{\@secondoftwo}%
\providecommand \bibfield  [0]{\@secondoftwo}%
\providecommand \translation [1]{[#1]}%
\providecommand \BibitemOpen [0]{}%
\providecommand \bibitemStop [0]{}%
\providecommand \bibitemNoStop [0]{.\EOS\space}%
\providecommand \EOS [0]{\spacefactor3000\relax}%
\providecommand \BibitemShut  [1]{\csname bibitem#1\endcsname}%
\let\auto@bib@innerbib\@empty
%</preamble>
\bibitem [{\citenamefont {Deng}\ \emph {et~al.}(2002)\citenamefont {Deng},
  \citenamefont {Weihs}, \citenamefont {Santori}, \citenamefont {Bloch},\ and\
  \citenamefont {Yamamoto}}]{deng_condensation_2002}%
  \BibitemOpen
  \bibfield  {author} {\bibinfo {author} {\bibfnamefont {H.}~\bibnamefont
  {Deng}}, \bibinfo {author} {\bibfnamefont {G.}~\bibnamefont {Weihs}},
  \bibinfo {author} {\bibfnamefont {C.}~\bibnamefont {Santori}}, \bibinfo
  {author} {\bibfnamefont {J.}~\bibnamefont {Bloch}},\ and\ \bibinfo {author}
  {\bibfnamefont {Y.}~\bibnamefont {Yamamoto}},\ }\bibfield  {title} {\bibinfo
  {title} {Condensation of {Semiconductor} {Microcavity} {Exciton}
  {Polaritons}},\ }\href {https://doi.org/10.1126/science.1074464} {\bibfield
  {journal} {\bibinfo  {journal} {Science}\ }\textbf {\bibinfo {volume}
  {298}},\ \bibinfo {pages} {199} (\bibinfo {year} {2002})}\BibitemShut
  {NoStop}%
\bibitem [{\citenamefont {Kasprzak}\ \emph {et~al.}(2006)\citenamefont
  {Kasprzak}, \citenamefont {Richard}, \citenamefont {Kundermann},
  \citenamefont {Baas}, \citenamefont {Jeambrun}, \citenamefont {Keeling},
  \citenamefont {Marchetti}, \citenamefont {Szymanska}, \citenamefont {Andre},
  \citenamefont {Staehli}, \citenamefont {Savona}, \citenamefont {Littlewood},
  \citenamefont {Deveaud},\ and\ \citenamefont
  {Dang}}]{kasprzak_bose-einstein_2006}%
  \BibitemOpen
  \bibfield  {author} {\bibinfo {author} {\bibfnamefont {J.}~\bibnamefont
  {Kasprzak}}, \bibinfo {author} {\bibfnamefont {M.}~\bibnamefont {Richard}},
  \bibinfo {author} {\bibfnamefont {S.}~\bibnamefont {Kundermann}}, \bibinfo
  {author} {\bibfnamefont {A.}~\bibnamefont {Baas}}, \bibinfo {author}
  {\bibfnamefont {P.}~\bibnamefont {Jeambrun}}, \bibinfo {author}
  {\bibfnamefont {J.~M.~J.}\ \bibnamefont {Keeling}}, \bibinfo {author}
  {\bibfnamefont {F.~M.}\ \bibnamefont {Marchetti}}, \bibinfo {author}
  {\bibfnamefont {M.~H.}\ \bibnamefont {Szymanska}}, \bibinfo {author}
  {\bibfnamefont {R.}~\bibnamefont {Andre}}, \bibinfo {author} {\bibfnamefont
  {J.~L.}\ \bibnamefont {Staehli}}, \bibinfo {author} {\bibfnamefont
  {V.}~\bibnamefont {Savona}}, \bibinfo {author} {\bibfnamefont {P.~B.}\
  \bibnamefont {Littlewood}}, \bibinfo {author} {\bibfnamefont
  {B.}~\bibnamefont {Deveaud}},\ and\ \bibinfo {author} {\bibfnamefont {L.~S.}\
  \bibnamefont {Dang}},\ }\bibfield  {title} {\bibinfo {title} {Bose-{Einstein}
  condensation of exciton polaritons},\ }\href
  {https://doi.org/10.1038/nature05131} {\bibfield  {journal} {\bibinfo
  {journal} {Nature}\ }\textbf {\bibinfo {volume} {443}},\ \bibinfo {pages}
  {409} (\bibinfo {year} {2006})}\BibitemShut {NoStop}%
\bibitem [{\citenamefont {Byrnes}\ \emph {et~al.}(2014)\citenamefont {Byrnes},
  \citenamefont {Kim},\ and\ \citenamefont
  {Yamamoto}}]{byrnes_excitonpolariton_2014}%
  \BibitemOpen
  \bibfield  {author} {\bibinfo {author} {\bibfnamefont {T.}~\bibnamefont
  {Byrnes}}, \bibinfo {author} {\bibfnamefont {N.~Y.}\ \bibnamefont {Kim}},\
  and\ \bibinfo {author} {\bibfnamefont {Y.}~\bibnamefont {Yamamoto}},\
  }\bibfield  {title} {\bibinfo {title} {Exciton-polariton condensates},\
  }\href {https://doi.org/10.1038/nphys3143} {\bibfield  {journal} {\bibinfo
  {journal} {Nature Physics}\ }\textbf {\bibinfo {volume} {10}},\ \bibinfo
  {pages} {803} (\bibinfo {year} {2014})}\BibitemShut {NoStop}%
\bibitem [{\citenamefont {Balili}\ \emph {et~al.}(2007)\citenamefont {Balili},
  \citenamefont {Hartwell}, \citenamefont {Snoke}, \citenamefont {Pfeiffer},\
  and\ \citenamefont {West}}]{balili_bose-einstein_2007}%
  \BibitemOpen
  \bibfield  {author} {\bibinfo {author} {\bibfnamefont {R.}~\bibnamefont
  {Balili}}, \bibinfo {author} {\bibfnamefont {V.}~\bibnamefont {Hartwell}},
  \bibinfo {author} {\bibfnamefont {D.}~\bibnamefont {Snoke}}, \bibinfo
  {author} {\bibfnamefont {L.}~\bibnamefont {Pfeiffer}},\ and\ \bibinfo
  {author} {\bibfnamefont {K.}~\bibnamefont {West}},\ }\bibfield  {title}
  {\bibinfo {title} {Bose-{Einstein} {Condensation} of {Microcavity}
  {Polaritons} in a {Trap}},\ }\href {https://doi.org/10.1126/science.1140990}
  {\bibfield  {journal} {\bibinfo  {journal} {Science}\ }\textbf {\bibinfo
  {volume} {316}},\ \bibinfo {pages} {1007} (\bibinfo {year}
  {2007})}\BibitemShut {NoStop}%
\bibitem [{\citenamefont {Plumhof}\ \emph {et~al.}(2014)\citenamefont
  {Plumhof}, \citenamefont {St{\"o}ferle}, \citenamefont {Mai}, \citenamefont
  {Scherf},\ and\ \citenamefont {Mahrt}}]{plumhof_room-temperature_2014}%
  \BibitemOpen
  \bibfield  {author} {\bibinfo {author} {\bibfnamefont {J.~D.}\ \bibnamefont
  {Plumhof}}, \bibinfo {author} {\bibfnamefont {T.}~\bibnamefont
  {St{\"o}ferle}}, \bibinfo {author} {\bibfnamefont {L.}~\bibnamefont {Mai}},
  \bibinfo {author} {\bibfnamefont {U.}~\bibnamefont {Scherf}},\ and\ \bibinfo
  {author} {\bibfnamefont {R.~F.}\ \bibnamefont {Mahrt}},\ }\bibfield  {title}
  {\bibinfo {title} {Room-temperature {Bose}-{Einstein} condensation of cavity
  exciton-polaritons in a polymer},\ }\href {https://doi.org/10.1038/nmat3825}
  {\bibfield  {journal} {\bibinfo  {journal} {Nature Materials}\ }\textbf
  {\bibinfo {volume} {13}},\ \bibinfo {pages} {247} (\bibinfo {year}
  {2014})}\BibitemShut {NoStop}%
\bibitem [{\citenamefont {Sun}\ \emph {et~al.}(2017)\citenamefont {Sun},
  \citenamefont {Wen}, \citenamefont {Yoon}, \citenamefont {Liu}, \citenamefont
  {Steger}, \citenamefont {Pfeiffer}, \citenamefont {West}, \citenamefont
  {Snoke},\ and\ \citenamefont {Nelson}}]{sun_bose-einstein_2017}%
  \BibitemOpen
  \bibfield  {author} {\bibinfo {author} {\bibfnamefont {Y.}~\bibnamefont
  {Sun}}, \bibinfo {author} {\bibfnamefont {P.}~\bibnamefont {Wen}}, \bibinfo
  {author} {\bibfnamefont {Y.}~\bibnamefont {Yoon}}, \bibinfo {author}
  {\bibfnamefont {G.}~\bibnamefont {Liu}}, \bibinfo {author} {\bibfnamefont
  {M.}~\bibnamefont {Steger}}, \bibinfo {author} {\bibfnamefont {L.~N.}\
  \bibnamefont {Pfeiffer}}, \bibinfo {author} {\bibfnamefont {K.}~\bibnamefont
  {West}}, \bibinfo {author} {\bibfnamefont {D.~W.}\ \bibnamefont {Snoke}},\
  and\ \bibinfo {author} {\bibfnamefont {K.~A.}\ \bibnamefont {Nelson}},\
  }\bibfield  {title} {\bibinfo {title} {Bose-{Einstein} {Condensation} of
  {Long}-{Lifetime} {Polaritons} in {Thermal} {Equilibrium}},\ }\href
  {https://doi.org/10.1103/PhysRevLett.118.016602} {\bibfield  {journal}
  {\bibinfo  {journal} {Phys. Rev. Lett.}\ }\textbf {\bibinfo {volume} {118}},\
  \bibinfo {pages} {016602} (\bibinfo {year} {2017})}\BibitemShut {NoStop}%
\bibitem [{\citenamefont {Sun}\ \emph {et~al.}(2012)\citenamefont {Sun},
  \citenamefont {Jia}, \citenamefont {Barsi}, \citenamefont {Rica},
  \citenamefont {Picozzi},\ and\ \citenamefont
  {Fleischer}}]{sun_observation_2012}%
  \BibitemOpen
  \bibfield  {author} {\bibinfo {author} {\bibfnamefont {C.}~\bibnamefont
  {Sun}}, \bibinfo {author} {\bibfnamefont {S.}~\bibnamefont {Jia}}, \bibinfo
  {author} {\bibfnamefont {C.}~\bibnamefont {Barsi}}, \bibinfo {author}
  {\bibfnamefont {S.}~\bibnamefont {Rica}}, \bibinfo {author} {\bibfnamefont
  {A.}~\bibnamefont {Picozzi}},\ and\ \bibinfo {author} {\bibfnamefont {J.~W.}\
  \bibnamefont {Fleischer}},\ }\bibfield  {title} {\bibinfo {title}
  {Observation of the kinetic condensation of classical waves},\ }\href
  {https://doi.org/10.1038/nphys2278} {\bibfield  {journal} {\bibinfo
  {journal} {Nature Physics}\ }\textbf {\bibinfo {volume} {8}},\ \bibinfo
  {pages} {470} (\bibinfo {year} {2012})}\BibitemShut {NoStop}%
\bibitem [{\citenamefont {Klaers}\ \emph
  {et~al.}(2010{\natexlab{a}})\citenamefont {Klaers}, \citenamefont
  {Vewinger},\ and\ \citenamefont {Weitz}}]{klaers_thermalization_2010}%
  \BibitemOpen
  \bibfield  {author} {\bibinfo {author} {\bibfnamefont {J.}~\bibnamefont
  {Klaers}}, \bibinfo {author} {\bibfnamefont {F.}~\bibnamefont {Vewinger}},\
  and\ \bibinfo {author} {\bibfnamefont {M.}~\bibnamefont {Weitz}},\ }\bibfield
   {title} {\bibinfo {title} {Thermalization of a two-dimensional photonic gas
  in a ‘white wall’ photon box},\ }\href
  {https://doi.org/10.1038/nphys1680} {\bibfield  {journal} {\bibinfo
  {journal} {Nature Physics}\ }\textbf {\bibinfo {volume} {6}},\ \bibinfo
  {pages} {512} (\bibinfo {year} {2010}{\natexlab{a}})}\BibitemShut {NoStop}%
\bibitem [{\citenamefont {Klaers}\ \emph
  {et~al.}(2010{\natexlab{b}})\citenamefont {Klaers}, \citenamefont {Schmitt},
  \citenamefont {Vewinger},\ and\ \citenamefont
  {Weitz}}]{klaers_boseeinstein_2010}%
  \BibitemOpen
  \bibfield  {author} {\bibinfo {author} {\bibfnamefont {J.}~\bibnamefont
  {Klaers}}, \bibinfo {author} {\bibfnamefont {J.}~\bibnamefont {Schmitt}},
  \bibinfo {author} {\bibfnamefont {F.}~\bibnamefont {Vewinger}},\ and\
  \bibinfo {author} {\bibfnamefont {M.}~\bibnamefont {Weitz}},\ }\bibfield
  {title} {\bibinfo {title} {Bose-{Einstein} condensation of photons in an
  optical microcavity},\ }\href {https://doi.org/10.1038/nature09567}
  {\bibfield  {journal} {\bibinfo  {journal} {Nature}\ }\textbf {\bibinfo
  {volume} {468}},\ \bibinfo {pages} {545} (\bibinfo {year}
  {2010}{\natexlab{b}})}\BibitemShut {NoStop}%
\bibitem [{\citenamefont {Marelic}\ and\ \citenamefont
  {Nyman}(2015)}]{marelic_experimental_2015}%
  \BibitemOpen
  \bibfield  {author} {\bibinfo {author} {\bibfnamefont {J.}~\bibnamefont
  {Marelic}}\ and\ \bibinfo {author} {\bibfnamefont {R.~A.}\ \bibnamefont
  {Nyman}},\ }\bibfield  {title} {\bibinfo {title} {Experimental evidence for
  inhomogeneous pumping and energy-dependent effects in photon
  {Bose}-{Einstein} condensation},\ }\href
  {https://doi.org/10.1103/PhysRevA.91.033813} {\bibfield  {journal} {\bibinfo
  {journal} {Phys. Rev. A}\ }\textbf {\bibinfo {volume} {91}},\ \bibinfo
  {pages} {033813} (\bibinfo {year} {2015})}\BibitemShut {NoStop}%
\bibitem [{\citenamefont {Weill}\ \emph {et~al.}(2019)\citenamefont {Weill},
  \citenamefont {Bekker}, \citenamefont {Levit},\ and\ \citenamefont
  {Fischer}}]{weill_boseeinstein_2019}%
  \BibitemOpen
  \bibfield  {author} {\bibinfo {author} {\bibfnamefont {R.}~\bibnamefont
  {Weill}}, \bibinfo {author} {\bibfnamefont {A.}~\bibnamefont {Bekker}},
  \bibinfo {author} {\bibfnamefont {B.}~\bibnamefont {Levit}},\ and\ \bibinfo
  {author} {\bibfnamefont {B.}~\bibnamefont {Fischer}},\ }\bibfield  {title}
  {\bibinfo {title} {Bose-{Einstein} condensation of photons in an
  erbium-ytterbium co-doped fiber cavity},\ }\href
  {https://doi.org/10.1038/s41467-019-08527-0} {\bibfield  {journal} {\bibinfo
  {journal} {Nature Communications}\ }\textbf {\bibinfo {volume} {10}},\
  \bibinfo {pages} {747} (\bibinfo {year} {2019})}\BibitemShut {NoStop}%
\bibitem [{\citenamefont {Rajan}\ \emph {et~al.}(2016)\citenamefont {Rajan},
  \citenamefont {Ramesh~Babu},\ and\ \citenamefont
  {Senthilnathan}}]{rajan_photon_2016}%
  \BibitemOpen
  \bibfield  {author} {\bibinfo {author} {\bibfnamefont {R.}~\bibnamefont
  {Rajan}}, \bibinfo {author} {\bibfnamefont {P.}~\bibnamefont {Ramesh~Babu}},\
  and\ \bibinfo {author} {\bibfnamefont {K.}~\bibnamefont {Senthilnathan}},\
  }\bibfield  {title} {\bibinfo {title} {Photon condensation: {A} new paradigm
  for {Bose}-{Einstein} condensation},\ }\href
  {https://doi.org/10.1007/s11467-016-0568-3} {\bibfield  {journal} {\bibinfo
  {journal} {Front. Phys.}\ }\textbf {\bibinfo {volume} {11}},\ \bibinfo
  {pages} {110502} (\bibinfo {year} {2016})}\BibitemShut {NoStop}%
\bibitem [{\citenamefont {Nyman}\ and\ \citenamefont
  {Walker}(2018)}]{nyman_bose-einstein_2018}%
  \BibitemOpen
  \bibfield  {author} {\bibinfo {author} {\bibfnamefont {R.~A.}\ \bibnamefont
  {Nyman}}\ and\ \bibinfo {author} {\bibfnamefont {B.~T.}\ \bibnamefont
  {Walker}},\ }\bibfield  {title} {\bibinfo {title} {Bose-{Einstein}
  condensation of photons from the thermodynamic limit to small photon
  numbers},\ }\href {https://doi.org/10.1080/09500340.2017.1404655} {\bibfield
  {journal} {\bibinfo  {journal} {Journal of Modern Optics}\ }\textbf {\bibinfo
  {volume} {65}},\ \bibinfo {pages} {754} (\bibinfo {year} {2018})}\BibitemShut
  {NoStop}%
\bibitem [{\citenamefont {Walker}\ \emph {et~al.}(2018)\citenamefont {Walker},
  \citenamefont {Flatten}, \citenamefont {Hesten}, \citenamefont {Mintert},
  \citenamefont {Hunger}, \citenamefont {Trichet}, \citenamefont {Smith},\ and\
  \citenamefont {Nyman}}]{walker_driven-dissipative_2018}%
  \BibitemOpen
  \bibfield  {author} {\bibinfo {author} {\bibfnamefont {B.~T.}\ \bibnamefont
  {Walker}}, \bibinfo {author} {\bibfnamefont {L.~C.}\ \bibnamefont {Flatten}},
  \bibinfo {author} {\bibfnamefont {H.~J.}\ \bibnamefont {Hesten}}, \bibinfo
  {author} {\bibfnamefont {F.}~\bibnamefont {Mintert}}, \bibinfo {author}
  {\bibfnamefont {D.}~\bibnamefont {Hunger}}, \bibinfo {author} {\bibfnamefont
  {A.~A.~P.}\ \bibnamefont {Trichet}}, \bibinfo {author} {\bibfnamefont
  {J.~M.}\ \bibnamefont {Smith}},\ and\ \bibinfo {author} {\bibfnamefont
  {R.~A.}\ \bibnamefont {Nyman}},\ }\bibfield  {title} {\bibinfo {title}
  {Driven-dissipative non-equilibrium {Bose}-{Einstein} condensation of less
  than ten photons},\ }\href {https://doi.org/10.1038/s41567-018-0270-1}
  {\bibfield  {journal} {\bibinfo  {journal} {Nature Physics}\ }\textbf
  {\bibinfo {volume} {14}},\ \bibinfo {pages} {1173} (\bibinfo {year}
  {2018})}\BibitemShut {NoStop}%
\bibitem [{\citenamefont {Marelic}\ \emph {et~al.}(2016)\citenamefont
  {Marelic}, \citenamefont {Zajiczek}, \citenamefont {Hesten}, \citenamefont
  {Leung}, \citenamefont {Ong}, \citenamefont {Mintert},\ and\ \citenamefont
  {Nyman}}]{marelic2016spatiotemporal}%
  \BibitemOpen
  \bibfield  {author} {\bibinfo {author} {\bibfnamefont {J.}~\bibnamefont
  {Marelic}}, \bibinfo {author} {\bibfnamefont {L.~F.}\ \bibnamefont
  {Zajiczek}}, \bibinfo {author} {\bibfnamefont {H.~J.}\ \bibnamefont
  {Hesten}}, \bibinfo {author} {\bibfnamefont {K.~H.}\ \bibnamefont {Leung}},
  \bibinfo {author} {\bibfnamefont {E.~Y.~X.}\ \bibnamefont {Ong}}, \bibinfo
  {author} {\bibfnamefont {F.}~\bibnamefont {Mintert}},\ and\ \bibinfo {author}
  {\bibfnamefont {R.~A.}\ \bibnamefont {Nyman}},\ }\bibfield  {title} {\bibinfo
  {title} {Spatiotemporal coherence of non-equilibrium multimode photon
  condensates},\ }\href@noop {} {\bibfield  {journal} {\bibinfo  {journal} {New
  Journal of Physics}\ }\textbf {\bibinfo {volume} {18}},\ \bibinfo {pages}
  {103012} (\bibinfo {year} {2016})}\BibitemShut {NoStop}%
\bibitem [{\citenamefont {Schmitt}\ \emph {et~al.}(2014)\citenamefont
  {Schmitt}, \citenamefont {Damm}, \citenamefont {Dung}, \citenamefont
  {Vewinger}, \citenamefont {Klaers},\ and\ \citenamefont
  {Weitz}}]{schmitt_observation_2014}%
  \BibitemOpen
  \bibfield  {author} {\bibinfo {author} {\bibfnamefont {J.}~\bibnamefont
  {Schmitt}}, \bibinfo {author} {\bibfnamefont {T.}~\bibnamefont {Damm}},
  \bibinfo {author} {\bibfnamefont {D.}~\bibnamefont {Dung}}, \bibinfo {author}
  {\bibfnamefont {F.}~\bibnamefont {Vewinger}}, \bibinfo {author}
  {\bibfnamefont {J.}~\bibnamefont {Klaers}},\ and\ \bibinfo {author}
  {\bibfnamefont {M.}~\bibnamefont {Weitz}},\ }\bibfield  {title} {\bibinfo
  {title} {Observation of {Grand}-{Canonical} {Number} {Statistics} in a
  {Photon} {Bose}-{Einstein} {Condensate}},\ }\href
  {https://doi.org/10.1103/PhysRevLett.112.030401} {\bibfield  {journal}
  {\bibinfo  {journal} {Phys. Rev. Lett.}\ }\textbf {\bibinfo {volume} {112}},\
  \bibinfo {pages} {030401} (\bibinfo {year} {2014})}\BibitemShut {NoStop}%
\bibitem [{\citenamefont {Keeling}\ and\ \citenamefont
  {Kirton}(2016)}]{keeling_spatial_2016}%
  \BibitemOpen
  \bibfield  {author} {\bibinfo {author} {\bibfnamefont {J.}~\bibnamefont
  {Keeling}}\ and\ \bibinfo {author} {\bibfnamefont {P.}~\bibnamefont
  {Kirton}},\ }\bibfield  {title} {\bibinfo {title} {Spatial dynamics,
  thermalization, and gain clamping in a photon condensate},\ }\href
  {https://doi.org/10.1103/PhysRevA.93.013829} {\bibfield  {journal} {\bibinfo
  {journal} {Phys. Rev. A}\ }\textbf {\bibinfo {volume} {93}},\ \bibinfo
  {pages} {013829} (\bibinfo {year} {2016})}\BibitemShut {NoStop}%
\bibitem [{\citenamefont {Kirton}\ and\ \citenamefont
  {Keeling}(2015)}]{kirton_thermalization_2015}%
  \BibitemOpen
  \bibfield  {author} {\bibinfo {author} {\bibfnamefont {P.}~\bibnamefont
  {Kirton}}\ and\ \bibinfo {author} {\bibfnamefont {J.}~\bibnamefont
  {Keeling}},\ }\bibfield  {title} {\bibinfo {title} {Thermalization and
  breakdown of thermalization in photon condensates},\ }\href
  {https://doi.org/10.1103/PhysRevA.91.033826} {\bibfield  {journal} {\bibinfo
  {journal} {Phys. Rev. A}\ }\textbf {\bibinfo {volume} {91}},\ \bibinfo
  {pages} {033826} (\bibinfo {year} {2015})}\BibitemShut {NoStop}%
\bibitem [{\citenamefont {Hesten}\ \emph
  {et~al.}(2018{\natexlab{a}})\citenamefont {Hesten}, \citenamefont {Walker},
  \citenamefont {Nyman},\ and\ \citenamefont
  {Mintert}}]{hesten_collective_2018}%
  \BibitemOpen
  \bibfield  {author} {\bibinfo {author} {\bibfnamefont {H.~J.}\ \bibnamefont
  {Hesten}}, \bibinfo {author} {\bibfnamefont {B.}~\bibnamefont {Walker}},
  \bibinfo {author} {\bibfnamefont {R.~A.}\ \bibnamefont {Nyman}},\ and\
  \bibinfo {author} {\bibfnamefont {F.}~\bibnamefont {Mintert}},\ }\bibfield
  {title} {\bibinfo {title} {Collective excitation profiles and the dynamics of
  photonic condensates},\ }\href {http://arxiv.org/abs/1809.08774} {\bibfield
  {journal} {\bibinfo  {journal} {arXiv:1809.08774 [quant-ph]}\ } (\bibinfo
  {year} {2018}{\natexlab{a}})}\BibitemShut {NoStop}%
\bibitem [{\citenamefont {Hesten}\ \emph
  {et~al.}(2018{\natexlab{b}})\citenamefont {Hesten}, \citenamefont {Walker},
  \citenamefont {Nyman},\ and\ \citenamefont
  {Mintert}}]{hesten_non-critical_2018}%
  \BibitemOpen
  \bibfield  {author} {\bibinfo {author} {\bibfnamefont {H.~J.}\ \bibnamefont
  {Hesten}}, \bibinfo {author} {\bibfnamefont {B.~T.}\ \bibnamefont {Walker}},
  \bibinfo {author} {\bibfnamefont {R.~A.}\ \bibnamefont {Nyman}},\ and\
  \bibinfo {author} {\bibfnamefont {F.}~\bibnamefont {Mintert}},\ }\bibfield
  {title} {\bibinfo {title} {Non-critical slowing down of photonic
  condensation},\ }\href {https://arxiv.org/abs/1809.08772} {\bibfield
  {journal} {\bibinfo  {journal} {arXiv:1809.08772 [quant-ph]}\ } (\bibinfo
  {year} {2018}{\natexlab{b}})}\BibitemShut {NoStop}%
\bibitem [{\citenamefont {Schmitt}(2018)}]{schmitt_dynamics_2018}%
  \BibitemOpen
  \bibfield  {author} {\bibinfo {author} {\bibfnamefont {J.}~\bibnamefont
  {Schmitt}},\ }\bibfield  {title} {\bibinfo {title} {Dynamics and correlations
  of a {Bose}-{Einstein} condensate of photons},\ }\href
  {https://doi.org/10.1088/1361-6455/aad409} {\bibfield  {journal} {\bibinfo
  {journal} {J. Phys. B: At. Mol. Opt. Phys.}\ }\textbf {\bibinfo {volume}
  {51}},\ \bibinfo {pages} {173001} (\bibinfo {year} {2018})}\BibitemShut
  {NoStop}%
\bibitem [{\citenamefont {Schmitt}\ \emph {et~al.}(2015)\citenamefont
  {Schmitt}, \citenamefont {Damm}, \citenamefont {Dung}, \citenamefont
  {Vewinger}, \citenamefont {Klaers},\ and\ \citenamefont
  {Weitz}}]{schmitt_thermalization_2015}%
  \BibitemOpen
  \bibfield  {author} {\bibinfo {author} {\bibfnamefont {J.}~\bibnamefont
  {Schmitt}}, \bibinfo {author} {\bibfnamefont {T.}~\bibnamefont {Damm}},
  \bibinfo {author} {\bibfnamefont {D.}~\bibnamefont {Dung}}, \bibinfo {author}
  {\bibfnamefont {F.}~\bibnamefont {Vewinger}}, \bibinfo {author}
  {\bibfnamefont {J.}~\bibnamefont {Klaers}},\ and\ \bibinfo {author}
  {\bibfnamefont {M.}~\bibnamefont {Weitz}},\ }\bibfield  {title} {\bibinfo
  {title} {Thermalization kinetics of light: {From} laser dynamics to
  equilibrium condensation of photons},\ }\href
  {https://doi.org/10.1103/PhysRevA.92.011602} {\bibfield  {journal} {\bibinfo
  {journal} {Phys. Rev. A}\ }\textbf {\bibinfo {volume} {92}},\ \bibinfo
  {pages} {011602(R)} (\bibinfo {year} {2015})}\BibitemShut {NoStop}%
\bibitem [{\citenamefont {Schmitt}\ \emph {et~al.}(2016)\citenamefont
  {Schmitt}, \citenamefont {Damm}, \citenamefont {Dung}, \citenamefont
  {Vewinger}, \citenamefont {Klaers},\ and\ \citenamefont
  {Weitz}}]{schmitt_bose-einstein_2016}%
  \BibitemOpen
  \bibfield  {author} {\bibinfo {author} {\bibfnamefont {J.}~\bibnamefont
  {Schmitt}}, \bibinfo {author} {\bibfnamefont {T.}~\bibnamefont {Damm}},
  \bibinfo {author} {\bibfnamefont {D.}~\bibnamefont {Dung}}, \bibinfo {author}
  {\bibfnamefont {F.}~\bibnamefont {Vewinger}}, \bibinfo {author}
  {\bibfnamefont {J.}~\bibnamefont {Klaers}},\ and\ \bibinfo {author}
  {\bibfnamefont {M.}~\bibnamefont {Weitz}},\ }\bibfield  {title} {\bibinfo
  {title} {Bose-{Einstein} {Condensation} of {Photons} versus {Lasing} and
  {Hanbury} {Brown}-{Twiss} {Measurements} with a {Condensate} of {Light}},\
  }\href {https://doi.org/10.1142/9789813200616_0008} {\bibfield  {journal}
  {\bibinfo  {journal} {Laser Spectroscopy}\ ,\ \bibinfo {pages} {85}}
  (\bibinfo {year} {2016})}\BibitemShut {NoStop}%
\bibitem [{\citenamefont {Leymann}\ \emph {et~al.}(2017)\citenamefont
  {Leymann}, \citenamefont {Vorberg}, \citenamefont {Lettau}, \citenamefont
  {Hopfmann}, \citenamefont {Schneider}, \citenamefont {Kamp}, \citenamefont
  {H{\"o}fling}, \citenamefont {Ketzmerick}, \citenamefont {Wiersig},
  \citenamefont {Reitzenstein},\ and\ \citenamefont
  {Eckardt}}]{leymann_pump-power-driven_2017}%
  \BibitemOpen
  \bibfield  {author} {\bibinfo {author} {\bibfnamefont {H.~A.~M.}\
  \bibnamefont {Leymann}}, \bibinfo {author} {\bibfnamefont {D.}~\bibnamefont
  {Vorberg}}, \bibinfo {author} {\bibfnamefont {T.}~\bibnamefont {Lettau}},
  \bibinfo {author} {\bibfnamefont {C.}~\bibnamefont {Hopfmann}}, \bibinfo
  {author} {\bibfnamefont {C.}~\bibnamefont {Schneider}}, \bibinfo {author}
  {\bibfnamefont {M.}~\bibnamefont {Kamp}}, \bibinfo {author} {\bibfnamefont
  {S.}~\bibnamefont {H{\"o}fling}}, \bibinfo {author} {\bibfnamefont
  {R.}~\bibnamefont {Ketzmerick}}, \bibinfo {author} {\bibfnamefont
  {J.}~\bibnamefont {Wiersig}}, \bibinfo {author} {\bibfnamefont
  {S.}~\bibnamefont {Reitzenstein}},\ and\ \bibinfo {author} {\bibfnamefont
  {A.}~\bibnamefont {Eckardt}},\ }\bibfield  {title} {\bibinfo {title}
  {Pump-power-driven mode switching in a microcavity device and its relation to
  bose-einstein condensation},\ }\href@noop {} {\bibfield  {journal} {\bibinfo
  {journal} {Physical Review X}\ }\textbf {\bibinfo {volume} {7}},\ \bibinfo
  {pages} {021045} (\bibinfo {year} {2017})}\BibitemShut {NoStop}%
\bibitem [{\citenamefont {Vorberg}\ \emph {et~al.}(2018)\citenamefont
  {Vorberg}, \citenamefont {Ketzmerick},\ and\ \citenamefont
  {Eckardt}}]{vorberg2018unified}%
  \BibitemOpen
  \bibfield  {author} {\bibinfo {author} {\bibfnamefont {D.}~\bibnamefont
  {Vorberg}}, \bibinfo {author} {\bibfnamefont {R.}~\bibnamefont
  {Ketzmerick}},\ and\ \bibinfo {author} {\bibfnamefont {A.}~\bibnamefont
  {Eckardt}},\ }\bibfield  {title} {\bibinfo {title} {Unified theory for
  excited-state, fragmented, and equilibriumlike bose condensation in pumped
  photonic many-body systems},\ }\href@noop {} {\bibfield  {journal} {\bibinfo
  {journal} {Physical Review A}\ }\textbf {\bibinfo {volume} {97}},\ \bibinfo
  {pages} {063621} (\bibinfo {year} {2018})}\BibitemShut {NoStop}%
\bibitem [{\citenamefont {Radonji{\'{c}}}\ \emph {et~al.}(2018)\citenamefont
  {Radonji{\'{c}}}, \citenamefont {Kopylov}, \citenamefont {Balaž},\ and\
  \citenamefont {Pelster}}]{radonjic_interplay_2018}%
  \BibitemOpen
  \bibfield  {author} {\bibinfo {author} {\bibfnamefont {M.}~\bibnamefont
  {Radonji{\'{c}}}}, \bibinfo {author} {\bibfnamefont {W.}~\bibnamefont
  {Kopylov}}, \bibinfo {author} {\bibfnamefont {A.}~\bibnamefont {Balaž}},\
  and\ \bibinfo {author} {\bibfnamefont {A.}~\bibnamefont {Pelster}},\
  }\bibfield  {title} {\bibinfo {title} {Interplay of coherent and dissipative
  dynamics in condensates of light},\ }\href
  {https://doi.org/10.1088/1367-2630/aac2a6} {\bibfield  {journal} {\bibinfo
  {journal} {New J. Phys.}\ }\textbf {\bibinfo {volume} {20}},\ \bibinfo
  {pages} {055014} (\bibinfo {year} {2018})}\BibitemShut {NoStop}%
\bibitem [{\citenamefont {Hesten}\ \emph
  {et~al.}(2018{\natexlab{c}})\citenamefont {Hesten}, \citenamefont {Nyman},\
  and\ \citenamefont {Mintert}}]{hesten_decondensation_2018}%
  \BibitemOpen
  \bibfield  {author} {\bibinfo {author} {\bibfnamefont {H.~J.}\ \bibnamefont
  {Hesten}}, \bibinfo {author} {\bibfnamefont {R.~A.}\ \bibnamefont {Nyman}},\
  and\ \bibinfo {author} {\bibfnamefont {F.}~\bibnamefont {Mintert}},\
  }\bibfield  {title} {\bibinfo {title} {Decondensation in {Nonequilibrium}
  {Photonic} {Condensates}: {When} {Less} {Is} {More}},\ }\href
  {https://doi.org/10.1103/PhysRevLett.120.040601} {\bibfield  {journal}
  {\bibinfo  {journal} {Phys. Rev. Lett.}\ }\textbf {\bibinfo {volume} {120}},\
  \bibinfo {pages} {040601} (\bibinfo {year} {2018}{\natexlab{c}})}\BibitemShut
  {NoStop}%
\bibitem [{\citenamefont {Vlaho}\ \emph {et~al.}(2019)\citenamefont {Vlaho},
  \citenamefont {Leymann}, \citenamefont {Vorberg},\ and\ \citenamefont
  {Eckardt}}]{vlaho2019controlled}%
  \BibitemOpen
  \bibfield  {author} {\bibinfo {author} {\bibfnamefont {M.}~\bibnamefont
  {Vlaho}}, \bibinfo {author} {\bibfnamefont {H.~A.~M.}\ \bibnamefont
  {Leymann}}, \bibinfo {author} {\bibfnamefont {D.}~\bibnamefont {Vorberg}},\
  and\ \bibinfo {author} {\bibfnamefont {A.}~\bibnamefont {Eckardt}},\
  }\bibfield  {title} {\bibinfo {title} {Controlled two-mode emission from the
  interplay of driving and thermalization in a dye-filled photonic cavity},\
  }\href@noop {} {\bibfield  {journal} {\bibinfo  {journal} {Physical Review
  Research}\ }\textbf {\bibinfo {volume} {1}},\ \bibinfo {pages} {033191}
  (\bibinfo {year} {2019})}\BibitemShut {NoStop}%
\bibitem [{\citenamefont {Kirton}\ and\ \citenamefont
  {Keeling}(2013)}]{kirton_nonequilibrium_2013}%
  \BibitemOpen
  \bibfield  {author} {\bibinfo {author} {\bibfnamefont {P.}~\bibnamefont
  {Kirton}}\ and\ \bibinfo {author} {\bibfnamefont {J.}~\bibnamefont
  {Keeling}},\ }\bibfield  {title} {\bibinfo {title} {Nonequilibrium {Model} of
  {Photon} {Condensation}},\ }\href
  {https://doi.org/10.1103/PhysRevLett.111.100404} {\bibfield  {journal}
  {\bibinfo  {journal} {Phys. Rev. Lett.}\ }\textbf {\bibinfo {volume} {111}},\
  \bibinfo {pages} {100404} (\bibinfo {year} {2013})}\BibitemShut {NoStop}%
\bibitem [{\citenamefont {McCumber}(1964)}]{mccumber1964einstein}%
  \BibitemOpen
  \bibfield  {author} {\bibinfo {author} {\bibfnamefont {D.}~\bibnamefont
  {McCumber}},\ }\bibfield  {title} {\bibinfo {title} {Einstein relations
  connecting broadband emission and absorption spectra},\ }\href@noop {}
  {\bibfield  {journal} {\bibinfo  {journal} {Physical Review}\ }\textbf
  {\bibinfo {volume} {136}},\ \bibinfo {pages} {A954} (\bibinfo {year}
  {1964})}\BibitemShut {NoStop}%
\bibitem [{\citenamefont {Moroshkin}\ \emph {et~al.}(2014)\citenamefont
  {Moroshkin}, \citenamefont {Weller}, \citenamefont {Sa{\ss}}, \citenamefont
  {Klaers},\ and\ \citenamefont {Weitz}}]{moroshkin2014kennard}%
  \BibitemOpen
  \bibfield  {author} {\bibinfo {author} {\bibfnamefont {P.}~\bibnamefont
  {Moroshkin}}, \bibinfo {author} {\bibfnamefont {L.}~\bibnamefont {Weller}},
  \bibinfo {author} {\bibfnamefont {A.}~\bibnamefont {Sa{\ss}}}, \bibinfo
  {author} {\bibfnamefont {J.}~\bibnamefont {Klaers}},\ and\ \bibinfo {author}
  {\bibfnamefont {M.}~\bibnamefont {Weitz}},\ }\bibfield  {title} {\bibinfo
  {title} {Kennard-stepanov relation connecting absorption and emission spectra
  in an atomic gas},\ }\href@noop {} {\bibfield  {journal} {\bibinfo  {journal}
  {Physical review letters}\ }\textbf {\bibinfo {volume} {113}},\ \bibinfo
  {pages} {063002} (\bibinfo {year} {2014})}\BibitemShut {NoStop}%
\bibitem [{Note1()}]{Note1}%
  \BibitemOpen
  \bibinfo {note} {The pump can also be modeled as a very wide Gaussian beam,
  to more accurately correspond to the experiment. This introduces only minor
  quantitative changes into the work presented here, without effecting any of
  the qualitative results.}\BibitemShut {Stop}%
\bibitem [{\citenamefont {Nyman}(2017)}]{nyman_absorption_2017}%
  \BibitemOpen
  \bibfield  {author} {\bibinfo {author} {\bibfnamefont {R.~A.}\ \bibnamefont
  {Nyman}},\ }\href {https://doi.org/10.5281/zenodo.569817} {\bibinfo {title}
  {Absorption and {Fluorescence} spectra of {Rhodamine} 6g}} (\bibinfo {year}
  {2017})\BibitemShut {NoStop}%
\bibitem [{\citenamefont {Vorberg}\ \emph {et~al.}(2013)\citenamefont
  {Vorberg}, \citenamefont {Wustmann}, \citenamefont {Ketzmerick},\ and\
  \citenamefont {Eckardt}}]{vorberg_generalized_2013}%
  \BibitemOpen
  \bibfield  {author} {\bibinfo {author} {\bibfnamefont {D.}~\bibnamefont
  {Vorberg}}, \bibinfo {author} {\bibfnamefont {W.}~\bibnamefont {Wustmann}},
  \bibinfo {author} {\bibfnamefont {R.}~\bibnamefont {Ketzmerick}},\ and\
  \bibinfo {author} {\bibfnamefont {A.}~\bibnamefont {Eckardt}},\ }\bibfield
  {title} {\bibinfo {title} {Generalized {Bose}-{Einstein} {Condensation} into
  {Multiple} {States} in {Driven}-{Dissipative} {Systems}},\ }\href
  {https://doi.org/10.1103/PhysRevLett.111.240405} {\bibfield  {journal}
  {\bibinfo  {journal} {Phys. Rev. Lett.}\ }\textbf {\bibinfo {volume} {111}},\
  \bibinfo {pages} {240405} (\bibinfo {year} {2013})}\BibitemShut {NoStop}%
\bibitem [{\citenamefont {Vorberg}\ \emph {et~al.}(2015)\citenamefont
  {Vorberg}, \citenamefont {Wustmann}, \citenamefont {Schomerus}, \citenamefont
  {Ketzmerick},\ and\ \citenamefont {Eckardt}}]{vorberg2015nonequilibrium}%
  \BibitemOpen
  \bibfield  {author} {\bibinfo {author} {\bibfnamefont {D.}~\bibnamefont
  {Vorberg}}, \bibinfo {author} {\bibfnamefont {W.}~\bibnamefont {Wustmann}},
  \bibinfo {author} {\bibfnamefont {H.}~\bibnamefont {Schomerus}}, \bibinfo
  {author} {\bibfnamefont {R.}~\bibnamefont {Ketzmerick}},\ and\ \bibinfo
  {author} {\bibfnamefont {A.}~\bibnamefont {Eckardt}},\ }\bibfield  {title}
  {\bibinfo {title} {Nonequilibrium steady states of ideal bosonic and
  fermionic quantum gases},\ }\href@noop {} {\bibfield  {journal} {\bibinfo
  {journal} {Physical Review E}\ }\textbf {\bibinfo {volume} {92}},\ \bibinfo
  {pages} {062119} (\bibinfo {year} {2015})}\BibitemShut {NoStop}%
\bibitem [{\citenamefont {Siegman}(1986)}]{siegman_lasers_1986}%
  \BibitemOpen
  \bibfield  {author} {\bibinfo {author} {\bibfnamefont {A.~E.}\ \bibnamefont
  {Siegman}},\ }\href@noop {} {\emph {\bibinfo {title} {Lasers}}}\ (\bibinfo
  {publisher} {University Science Books},\ \bibinfo {address} {Mill Valley,
  Calif.},\ \bibinfo {year} {1986})\BibitemShut {NoStop}%
\bibitem [{\citenamefont {Klaers}\ \emph {et~al.}(2012)\citenamefont {Klaers},
  \citenamefont {Schmitt}, \citenamefont {Damm}, \citenamefont {Vewinger},\
  and\ \citenamefont {Weitz}}]{klaers2012statistical}%
  \BibitemOpen
  \bibfield  {author} {\bibinfo {author} {\bibfnamefont {J.}~\bibnamefont
  {Klaers}}, \bibinfo {author} {\bibfnamefont {J.}~\bibnamefont {Schmitt}},
  \bibinfo {author} {\bibfnamefont {T.}~\bibnamefont {Damm}}, \bibinfo {author}
  {\bibfnamefont {F.}~\bibnamefont {Vewinger}},\ and\ \bibinfo {author}
  {\bibfnamefont {M.}~\bibnamefont {Weitz}},\ }\bibfield  {title} {\bibinfo
  {title} {Statistical physics of bose-einstein-condensed light in a dye
  microcavity},\ }\href@noop {} {\bibfield  {journal} {\bibinfo  {journal}
  {Physical review letters}\ }\textbf {\bibinfo {volume} {108}},\ \bibinfo
  {pages} {160403} (\bibinfo {year} {2012})}\BibitemShut {NoStop}%
\bibitem [{\citenamefont {Dung}\ \emph {et~al.}(2017)\citenamefont {Dung},
  \citenamefont {Kurtscheid}, \citenamefont {Damm}, \citenamefont {Schmitt},
  \citenamefont {Vewinger}, \citenamefont {Weitz},\ and\ \citenamefont
  {~}}]{dung_variable_2017}%
  \BibitemOpen
  \bibfield  {author} {\bibinfo {author} {\bibfnamefont {D.}~\bibnamefont
  {Dung}}, \bibinfo {author} {\bibfnamefont {C.}~\bibnamefont {Kurtscheid}},
  \bibinfo {author} {\bibfnamefont {T.}~\bibnamefont {Damm}}, \bibinfo {author}
  {\bibfnamefont {J.}~\bibnamefont {Schmitt}}, \bibinfo {author} {\bibfnamefont
  {F.}~\bibnamefont {Vewinger}}, \bibinfo {author} {\bibfnamefont
  {M.}~\bibnamefont {Weitz}},\ and\ \bibinfo {author} {\bibfnamefont
  {J.}~\bibnamefont {~}},\ }\bibfield  {title} {\bibinfo {title} {Variable
  potentials for thermalized light and coupled condensates},\ }\href
  {https://doi.org/10.1038/nphoton.2017.139} {\bibfield  {journal} {\bibinfo
  {journal} {Nature Photonics}\ }\textbf {\bibinfo {volume} {11}},\ \bibinfo
  {pages} {565} (\bibinfo {year} {2017})}\BibitemShut {NoStop}%
\bibitem [{Note2()}]{Note2}%
  \BibitemOpen
  \bibinfo {note} {Alternatively, we could consider (a), (b) and (c) to
  correspond to different cavities, which have the same photon loss rate $
  \kappa $ at the given cutoff frequencies.}\BibitemShut {Stop}%
\end{thebibliography}%

\end{document}